\documentstyle[prl,aps,epsf]{revtex}
\def\be{\begin{equation}}
\def\ee{\end{equation}}
\def\ba{\begin{eqnarray}}
\def\ea{\end{eqnarray}}

\begin{document}
\draft
\title{
A new approach to the ground state of quantum-Hall systems. \\ Basic principles}

\author{
S. A. Mikhailov\cite{address} }

\address
{Institute for Theoretical Physics, University of Regensburg, 93040 Regensburg, Germany}
\date{\today}
\maketitle
\begin{abstract}
I present a new approach to the many-body ground state of quantum-Hall systems. The method describes the behavior of a two-dimensional electron system at all Landau-level filling factors $\nu$, continuously as a function of magnetic field, and enables one to analytically calculate any physical value, characterizing the system, at all $\nu$. New proposed trial many-body wave functions have a clear physical meaning and a very large variational freedom, which opens up wide possibilities to search for the ground state of the system at all magnetic fields.
\end{abstract}

\nopagebreak










\section{Introduction}
\label{intro}

The nature of the ground state of a two-dimensional (2D) electron system (ES) in a strong perpendicular magnetic field ${\bf B}=(0,0,B)$ was a subject of intensive theoretical investigations during the past years \cite{Fukuyama79,Yoshioka79,Bychkov81,Yoshioka83b,Maki83,Yoshioka83a,Laughlin83,Haldane83,Yoshioka84b,Lam84,Levesque84,Haldane85,MacDonald85,Kivelson86,Morf86,Jain89a,Halperin93,Aleiner95,Koulakov96,Fogler96,Moessner96,Fogler97,Jungwirth00,MacDonald00}. A great interest to this problem was motivated by the need to explain the integer \cite{Klitzing80} and the fractional \cite{Tsui82} quantum Hall effects, as well as to understand such fundamental physical phenomena as electron-electron correlations at a degenerate lowest Landau level and Wigner crystallization \cite{Wigner34} of 2D electrons in strong magnetic fields. Recently discovered novel intriguing peculiarities of 2D electron transport at higher Landau levels \cite{Lilly99a,Lilly99b,Du99,Pan99,Cooper99,Du00,Eisenstein00} provided further impetus to these efforts.

The problem of interacting 2D electrons in a magnetic field was discussed in the literature from different points of view. First of all, there exists a solution of the {\it classical many-electron} problem \cite{Chaplik72,Meissner76,Bonsall77}. If electrons are considered as classical point particles, they form, due to the Coulomb repulsion, a regular Wigner-crystal lattice \cite{Wigner34} of a triangular,

\be
{\bf R}_\triangle \sim a\left(l_1+l_2/2,\sqrt{3}l_2/2\right),{\ \ }n_sa^2=2/\sqrt{3},
\label{tr-lattice}
\ee
or a square symmetry,
\be
{\bf R}_\Box \sim a(l_1,l_2),{\ \ }n_sa^2=1.
\label{sq-lattice}
\ee
Here $l_1$ and $l_2$ are integer, and the lattice constant $a$ is related to the {\it average} electron density $n_s$. The energy (per particle) of the triangular electron lattice on a neutralizing positive background with the charge density $+|e|n_s$ is slightly lower than that of the square one,\cite{Meissner76,Bonsall77}

\be
E_\triangle^{WC,classical} = -1.960517, {\ \ } E_\Box^{WC,classical} = - 1.9501325
\label{latt-energy}
\ee
(throughout the paper I will use the $B$-independent energy units $e^2\sqrt{n_s}/\kappa$, where $e$ is the electron charge and $\kappa$ is the dielectric constant). The classical result (\ref{latt-energy}) does not depend on magnetic field.

Then, there exists the solution of a {\it quantum-mechanical single-electron} problem \cite{Fock28,Landau30,Teller30}. The single-particle Schr\"odinger equation 
\be
\frac 1{2m^\star}\left({\bf \hat p}+\frac {|e|}{2c}{\bf B}\times{\bf
r}\right)^2\phi({\bf r})=\epsilon\phi({\bf r}),
\label{1particle}
\ee 
(I use a symmetric gauge) has the solutions \cite{Fock28}

\be
\epsilon=\epsilon_{n,l}=\hbar \omega_c\left(n+\frac{l+|l|+1}2\right),{\ \ }0\le n<\infty, {\ \ }-\infty<l<+\infty,
\label{1particleenergy}
\ee

\be
\phi({\bf
r})=\phi_{n,l}({\bf
r})=\frac{e^{il\theta}}{\sqrt{\pi}\lambda}\left(\frac{n!}{(n+|l|)!}\right)^{1/2}\exp\left(-\frac{r^2}{2\lambda^2}\right)\left(\frac
r\lambda\right)^{|l|}L_n^{|l|}\left(\frac{r^2}{\lambda^2}\right).
\label{1particlefunctions}
\ee
Here $\hbar$, $c$, and $m^\star$ are the Planck constant, velocity of light, and the electron effective mass, respectively, $n$ and $l$ are integer, $\omega_c=|e|B/m^\star c$ is the cyclotron frequency, $\lambda^2=2l_B^2=2\hbar/m^\star\omega_c$, $l_B$ is the magnetic length, and $L_n^{|l|}$ are associated Laguerre polynomials. The states with $n=0$ and $-\infty<l\le 0$ belong to the lowest Landau level with the energy $\hbar \omega_c/2$ and the wave functions which I will write in the form

\be
\psi_L({\bf r})\equiv \phi_{0,l\le 0}({\bf r})= \frac{(z^*)^L}{\lambda\sqrt{\pi L!}}\exp(-zz^*/2),
\label{psiLfunc1}
\ee
where $L=-l= 0,1,2,\dots$ and $z=(x+iy)/\lambda$ is a complex dimensionless coordinate of an electron. 

The {\it quantum-mechanical many-electron} problem does not have a complete solution. As the problem is of extreme complexity, the main theoretical efforts were concentrated on attempts {\it to guess} the ground state many-body wave function, at separate points (or in certain intervals) of the magnetic field axis. For a completely filled lowest Landau level (the filling factor $\nu=\pi\lambda^2n_s=1$) a trial Hartree-Fock many-body wave function 

\be
\Psi_{HF}^{[N]}= \frac 1{\sqrt{N!}}\det_N|\psi_{L_j}({\bf r}_i)|,{\ \ }L_j=0,1,\dots,N-1,
\label{HF}
\ee
was proposed in Ref. \cite{Bychkov81} ($\det_N$ will denote the determinant of an $N\times N$ matrix). It has the form of a Slater determinant built from the lowest-Landau-level single-particle states (\ref{psiLfunc1}), and implies that in the $N$-electron system the states $L=0$ to $L=N-1$ are occupied by electrons. In the thermodynamic limit $N\to\infty$ the state (\ref{HF}) is characterized by the uniform electron density \cite{Bychkov81} $n_e({\bf r})=1/\pi\lambda^2$ and the energy per particle \cite{Laughlin83} (in units $e^2\sqrt{n_s}/\kappa$)

\be
\epsilon_{HF}=-\pi/2=-1.57080.
\label{HFenergy}
\ee
The Hartree-Fock wave function can describe the properties of the system only at the one point ($\nu=1$) of the $B$-axis; otherwise the density of electrons does not coincide with the density of the positive background, and the electro-neutrality requirement is violated. 

At a partial Landau-level filling $\nu<1$ one could expect an inhomogeneous distribution of the electron density. Therefore, first suggestions for the ground state of the 2DES at $\nu<1$ were related with the charge-density-wave \cite{Fukuyama79,Yoshioka79,Yoshioka83b}, or the Wigner crystal (WC) model \cite{Maki83}. The WC many-body wave function \cite{Maki83} 

\be
\Psi_{WC}^{[N]}=\frac 1{\sqrt{N!}}\det_N|\chi_{L=0}({\bf r}_i,{\bf R}_j)|,
\label{WC}
\ee
\be
\chi_L({\bf r}_i,{\bf R}_j)\equiv\chi_{ij}^L=\psi_L({\bf r}_i-{\bf R}_j) e^{-i\pi{\bf r}_i\cdot({\bf B\times R}_j)/\phi_0},
\label{chi-functions}
\ee
is a natural generalization of the classical Wigner-crystal solution: electrons are described by the gaussians $\psi_{L=0}$ centered at the points ${\bf R}_j$ of the triangular lattice (\ref{tr-lattice}) (here $\phi_0$ is the flux quantum). Contrary to the Hartree-Fock solution, the wave function (\ref{WC}) can describe the properties of the system at all $\nu$, continuously as a function of magnetic field, as the electro-neutrality requirement is automatically fulfilled due to the relevant choice of the lattice constant 
in Eqs. (\ref{tr-lattice})--(\ref{sq-lattice}). Due to intrinsic triangular symmetry of the solution (\ref{WC}), one could also expect \cite{Tosatti83,Maki83} that the $B$-dependence of the energy of the system contains oscillating functions of 
\be
\frac{BS_\triangle}{\phi_0}\propto\frac{B}{n_s\phi_0}=\frac 1\nu,
\ee
where $S_\triangle$ is the area of any triangle in the lattice (\ref{tr-lattice}). This could lead to certain peculiarities in the energy at $\nu=1/3$, 1/5, and other fractions. Detailed calculations made by Maki and Zotos \cite{Maki83} revealed however no commensuration features in the energy (an interesting modification of the model, with less than one electron per elementary cell, gave commensuration features at fractional $\nu$, but at a slightly higher energy \cite{Claro85,Claro87}). The reason for that is a small overlap of the single-particle gaussians $\psi_{L=0}$ centered at neighbor lattice points ${\bf R}_j$: in order to get the Aharonov-Bohm-type oscillations in the energy, electrons should be able to ``tunnel'' between different lattice points ${\bf R}_j$, but this is hampered by a negligible overlap of gaussians $\psi_{L=0}$, centered at different lattice points, at $\nu\le 1/3$. 

The trial many-body wave functions proposed by Laughlin \cite{Laughlin83}, 

\be
\Psi_{Laughlin}^{\nu=1/m}=\prod_{ij}(z_i-z_j)^m\prod_k e^{-|z_k|^2/2},
\label{Laughlin}
\ee
describe the system at fractional filling factors $\nu=1/m$, where $m$ is odd integer, and are characterized by the uniform 2D electron density $n_s=1/m\pi\lambda^2$. Laughlin showed that at $\nu=1/3$ and 1/5 the energy of the liquid state (\ref{Laughlin}) is {\it lower than} the energy of the Wigner crystal. This result was confirmed by exact-diagonalization calculations of the energy of a few-electron system \cite{Yoshioka83a,Haldane83,Yoshioka84b}. At $\nu=1$ the Laughlin wave function coincides with the Hartree-Fock one (\ref{HF}) and has the energy given by (\ref{HFenergy}). 

Subsequent theoretical work \cite{Haldane83,Lam84,Levesque84,Haldane85,MacDonald85,Kivelson86,Morf86,Jain89a,Halperin93} in the region $\nu<1$ was devoted to the development and further generalizations of the Laughlin theory to other fractional filling factors. It was established that the Laughlin liquid has the energy lower than the Wigner crystal at $\nu\gtrsim 1/7$; at $\nu\to 0$ the Wigner crystal state has the lower energy. 

In 1996 an interest aroused to the ground state properties at higher Landau levels $\nu\gg 1$. It was shown \cite{Koulakov96} that a unidirectional charge-density-wave state is likely the ground state of a 2DES at higher Landau levels. This conclusion has been confirmed in a number of subsequent publications \cite{Fogler96,Moessner96,Fogler97,Jungwirth00,MacDonald00} and turned out to be in agreement with recent experiments \cite{Lilly99a,Lilly99b,Du99,Pan99,Cooper99,Du00,Eisenstein00}.

The Laughlin-based approach describes the ground state properties of a 2DES at separate, specific values of the Landau-level filling factor. In this paper I present a new method, which describes the physical properties of the system at all $\nu$, {\it continuously} as a function of magnetic field. I discuss basic principles of the new method, with a short demonstration of its advantages, by an example of a few specific calculations. In Section \ref{sec:wavefunc} I discuss possible ways of solving the many-body Schr\"odinger equation and construct an infinite set of trial many-body wave functions for the ground state of the system. In Section \ref{sec:properties} I present an exact analytical method of calculating different physical properties of the proposed many-particle states, and show a few examples of calculated characteristics of the system. Conclusive remarks are given in Section \ref{sec:disc}. Mathematical details can be found in Appendices.

\section{Trial solutions of the many-body Schr\"odinger equation}
\label{sec:wavefunc}

\subsection{The problem}
\label{formulation}

I begin with the many-body Schr\"odinger equation for a system of $N$ 2D electrons, moving in the plane $z=0$ in a perpendicular magnetic field ${\bf B}=(0,0,B)$,
\be
\hat H \Psi^{[N]}({\bf r}_1,{\bf r}_2,\dots,{\bf r}_N) = E \Psi^{[N]}({\bf r}_1,{\bf r}_2,\dots,{\bf r}_N).
\label{SE}
\ee
The Hamiltonian $\hat H = \hat K + V_{ee} + V_{eb} + V_{bb}$ of the system consists of the kinetic energy term,

\be
\hat K =\sum_{i=1}^N \hat k_i=\frac 1{2m^\star}\sum_{i=1}^N \left(\hat {\bf p}_i+\frac {|e|}{2c}{\bf B}\times{\bf r}_i\right)^2,
\label{kinetic}
\ee
the energy of Coulomb interaction of 2D electrons with each other,

\be
V_{ee}=\frac 12\sum_{i=1}^N\sum_{j=1,j\neq i}^N \frac{e^2}{\kappa|{\bf r}_i-{\bf r}_j|},
\label{Vee}
\ee
the interaction energy of electrons with the positive background $V_{eb}$, and of the positive background with itself $V_{bb}$. Specific expressions for the terms $V_{eb}+V_{bb}$ depend on the model of the positive background (e.g. randomly distributed ionized donors or a uniformly charged jellium disk). The influence of disorder will only briefly be discussed in subsequent Sections. In the main part of the paper I will assume that the positive background has a form of a disk with a large radius $R_b$ and a uniform 2D charge density $+|e|n_s$. Electro-neutrality requires that the radius $R_b$ of the disk and the number of electrons in the system $N$ be related by the formula
\be 
\pi R_b^2n_s=N. 
\label{radiusR}
\ee
Then the energy associated with the positive background is written as

\be
V_{eb} + V_{bb}=- \int d{\bf r}\sum_{i=1}^N \frac{e^2n_b({\bf r})}{\kappa|{\bf r}_i-{\bf r}|}+ \frac {1}2 \int d{\bf r}\int d{\bf r}^\prime \frac {e^2n_b({\bf r})n_b({\bf r}^\prime)}{\kappa|{\bf r}-{\bf r}^\prime|},
\label{Vbb}
\ee
where $n_b({\bf r})=n_s\theta(R_b-r)$. I will assume that the magnetic field is sufficiently strong ($\nu\le 1$), so that all electron spins are aligned along the $z$-axis, and will search for an antisymmetric solution of the Schr\"odinger equation (\ref{SE}).

The many-body wave function $\Psi^{[N]}({\bf r}_1,{\bf r}_2,\dots,{\bf r}_N)$ is a function of $N$ vector variables, so that one needs a suitable set of basis functions in a $2N$-dimensional space. As in the literature there was an extensive discussion on the availability and completeness of different sets of functions in the considered problem \cite{Brown64,Dana83,Thouless84,Imai90,Ishikawa92,Ishikawa95,Ishikawa98,Ishikawa99,Maeda00}, I briefly discuss this point here. The basis set which I will use in this work consists of the eigenfunctions of the kinetic energy operator (\ref{kinetic}),
\be
\phi_{n_1l_1}({\bf r}_1)
\phi_{n_2l_2}({\bf r}_2)
\dots
\phi_{n_Nl_N}({\bf r}_N),{\ \ \ \ }0\le n_i<\infty, {\ \ }-\infty<l_i<+\infty.
\label{fullset}
\ee
This set is complete and orthonormal in a $2N$-dimensional space. A properly antisymmetrized subset of these functions,
\be
\frac{1}{\sqrt{N!}}\det_N
\left |\begin{array}{cccc}
\psi_{L_1}({\bf r}_1) & \psi_{L_2}({\bf r}_1) &
\dots & \psi_{L_N}({\bf r}_1) \\
\psi_{L_1}({\bf r}_2) & \psi_{L_2}({\bf r}_2) &
\dots & \psi_{L_N}({\bf r}_2) \\
\dots & \dots & \dots & \dots \\
\psi_{L_1}({\bf r}_N) & \psi_{L_2}({\bf r}_N) &
\dots & \psi_{L_N}({\bf r}_N)
\end{array}
\right |, {\ \ } 0\le L_1<L_2<\dots <L_N\le\infty,
\label{subset}
\ee
will be used for expansion of the {\it ground state} trial wave functions. The functions (\ref{subset}) are orthonormal in a $2N$-dimensional space and are the eigenfunctions of the kinetic energy operator $\hat K$ with the eigenenergy $N\hbar\omega_c/2$.

Now, I discuss two possible approaches to the problem (\ref{SE}).

\subsection{Perturbative approach}
\label{perturbative}

One could try to solve the problem (\ref{SE}) using the perturbative approach. Assume that the Coulomb energy of the system ($V_{ee} + V_{eb} + V_{bb}$) can be considered as a perturbation, and the non-perturbed Hamiltonian is given by the kinetic energy operator $\hat K$. The ground state of the non-perturbed problem is highly degenerate, so that one needs to use the degenerate perturbation theory. First, consider the Schr\"odinger equation (\ref{SE}) with $N=1$. This is a problem of a single 2D electron in the field of a disk-shaped ``impurity'' with the charge $+|e|$ and the radius $R_b=(\pi n_s)^{-1/2}$. According to the degenerate perturbation theory the ground-state wave function of the single-electron problem should be searched for in the form

\be
\Psi^{[N=1]}({\bf r})=\sum_{L=0}^\infty C_L\psi_L({\bf r}),
\label{1el-func}
\ee
with arbitrary coefficients $C_L$. Substitution of the wave function (\ref{1el-func}) into the Schr\"odinger equation (\ref{SE}) leads to a secular equation, which allows one to find, in principle, the energy and the wave function of the ground state of the one-electron problem. 

Now, consider two electrons moving in the field of a positively charged disk with the radius $R_b=(2/\pi n_s)^{1/2}$ and the total charge $+2|e|$. An antisymmetric ground-state two-electron wave function is expanded in the orthonormal set (\ref{subset}) with $N=2$,

\be
\Psi^{[N=2]}({\bf r}_1,{\bf r}_2)
= \sum_{L_1=0}^\infty \sum_{L_2>L_1}^\infty C_{L_1,L_2}\frac 1{\sqrt{2!}}\det_2\left |\begin{array}{cc}
\psi_{L_1}({\bf r}_1)& 
\psi_{L_2}({\bf r}_1)\\
\psi_{L_1}({\bf r}_2)& 
\psi_{L_2}({\bf r}_2)
\end{array}
\right |;
\label{2el-func}
\ee
here $C_{L_1,L_2}$ are arbitrary coefficients.
Substitution of the wave function (\ref{2el-func}) into the Schr\"odinger equation (\ref{SE}) leads to a secular equation, which gives, in principle, the energy and the wave function of the ground state of the two-electron problem. 

Similarly, in the problem of $N$ electrons moving in the field of a disk with the radius $R_b=(N/\pi n_s)^{1/2}$, Eq. (\ref{radiusR}), and the charge density $+|e|n_s$, the ground-state antisymmetric wave function should be searched for in the form
\be
\Psi^{[N]}({\bf r}_1,{\bf r}_2,\dots,{\bf r}_N) = \sum^\infty_{0\le L_1<L_2<\dots<L_N} C_{L_1,L_2,\dots,L_N}\frac{1}{\sqrt{N!}}\det_N
\left |
\psi_{L_j}({\bf r}_i)
\right |,
\label{Nel-func}
\ee
with arbitrary coefficients $C_{L_1,L_2,\dots,L_N}$.
The ground state of the system can be found, in principle, from the secular equation resulting from the substitution of the expansion (\ref{Nel-func}) into the Schr\"odinger equation (\ref{SE}). 

\subsection{Variational approach. Construction of trial many-body wave functions}
\label{variational}

Of course, it is absolutely impractical to try to solve the secular equations discussed above at $N\gg 1$. Instead, one can use the variational approach. Any trial many-body wave function in the $N$-electron problem can be searched for in the form of an arbitrary linear combination (\ref{Nel-func}). One of the possible trial wave functions is the Hartree-Fock solution (\ref{HF}). It corresponds to the trivial case 
\be
C_{L_1,L_2,\dots,L_N}^{HF}=\prod_{i=1}^N \delta_{L_i,i-1},
\label{HFcoefs}
\ee
when the coefficient $C_{L_1,L_2,\dots,L_N}$ vanishes at all but one points of the $N$-dimensional space of quantum numbers $\{L_1,L_2,\dots,L_N\}$. 
The number of other possible trial wave functions at all $\nu$ is however infinite, so that a simple physical idea would be desirable in order to narrow the number of possible linear combinations in Eq. (\ref{Nel-func}) down to a reasonable value. 

Such a simple physical idea has been actually used in the literature \cite{Maki83}. Consider the Wigner crystal wave function (\ref{WC}). It satisfies the electro-neutrality requirement at all magnetic fields, has a clear physical meaning and corresponds, in the limit of very strong magnetic fields $\nu\to 0$, to the classical picture of distribution of point particles in the triangular lattice. One can easily show that the wave function (\ref{WC}) can be explicitly expanded in a series of the type (\ref{Nel-func}). For the function $\chi_0({\bf r},{\bf R})$ I write
\be 
\chi_0({\bf r},{\bf R})\equiv\frac 1{\sqrt{\pi}\lambda} e^{-zz^*/2-ZZ^*/2+z^*Z}= \frac {e^{-zz^*/2-ZZ^*/2}}{\sqrt{\pi}\lambda}\sum_{L=0}^\infty \frac{(z^*Z)^L}{L!}= \sum_{L=0}^\infty A_L(Z)\psi_L({\bf r}),
\label{chi0expansion}
\ee
where $Z=(X+iY)/\lambda$, and 
\be
A_L(Z)= \frac{Z^L}{\sqrt{L!}}e^{-ZZ^*/2}.
\label{expanA}
\ee 

Then, the two-electron function (\ref{WC}) is presented as 
\ba
\Psi_{WC}^{[N=2]}
&=& \frac 1{\sqrt{2!}}
\det_2
\left |\begin{array}{cc}
\chi_0({\bf r}_1,{\bf R}_1)&
\chi_0({\bf r}_1,{\bf R}_2)\\
\chi_0({\bf r}_2,{\bf R}_1)&
\chi_0({\bf r}_2,{\bf R}_2)
\end{array}
\right |\nonumber \\ 
&=&\sum_{L_1=0}^\infty \sum_{L_2=0}^\infty A_{L_1}(Z_1)A_{L_2}(Z_2)\frac 1{\sqrt{2!}}\det_2\left |\begin{array}{cc}
\psi_{L_1}({\bf r}_1)& 
\psi_{L_2}({\bf r}_1)\\
\psi_{L_1}({\bf r}_2)& 
\psi_{L_2}({\bf r}_2)
\end{array}
\right |\nonumber \\ 
&=&\sum_{L_1=0}^\infty \sum_{L_2>L_1}^\infty \det_2
\left |\begin{array}{cc}
A_{L_1}(Z_1)&
A_{L_2}(Z_1)\\
A_{L_1}(Z_2)&
A_{L_2}(Z_2)
\end{array}
\right | \frac 1{\sqrt{2!}}\det_2
\left |\begin{array}{cc}
\psi_{L_1}({\bf r}_1)& 
\psi_{L_2}({\bf r}_1)\\
\psi_{L_1}({\bf r}_2)& 
\psi_{L_2}({\bf r}_2)
\end{array}
\right | .
\label{WC2expan}
\ea
Quite similarly, for the $N$-electron Wigner crystal wave function I obtain
\be
\Psi_{WC}^{[N]}
= \sum^\infty_{0\le L_1<L_2<\dots <L_N} \det_N
\left |A_{L_j}(Z_i)\right | \frac 1{\sqrt{N!}}\det_N
\left |\psi_{L_j}({\bf r}_i)
\right |.
\label{WCexpansion}
\ee
Eq. (\ref{WCexpansion}) explicitly gives the expansion coefficients $C_{L_1,L_2,\dots,L_N}$ of the Wigner crystal wave function (\ref{WC}) in the series of orthonormal functions (\ref{subset}),
\be
C_{L_1,L_2,\dots,L_N}^{WC}=\det_N \left |A_{L_j}(Z_i)\right |;
\label{WCcoefs}
\ee
the coefficients $C_{L_1,L_2,\dots,L_N}^{WC}$ are non-zero at almost all points of the $N$-dimensional space of quantum numbers $\{L_1,L_2,\dots,L_N\}$.

\begin{figure}
\begin{center}
\begin{math}
\begin{array}{ccc}
\epsfxsize=3.2cm
\epsffile{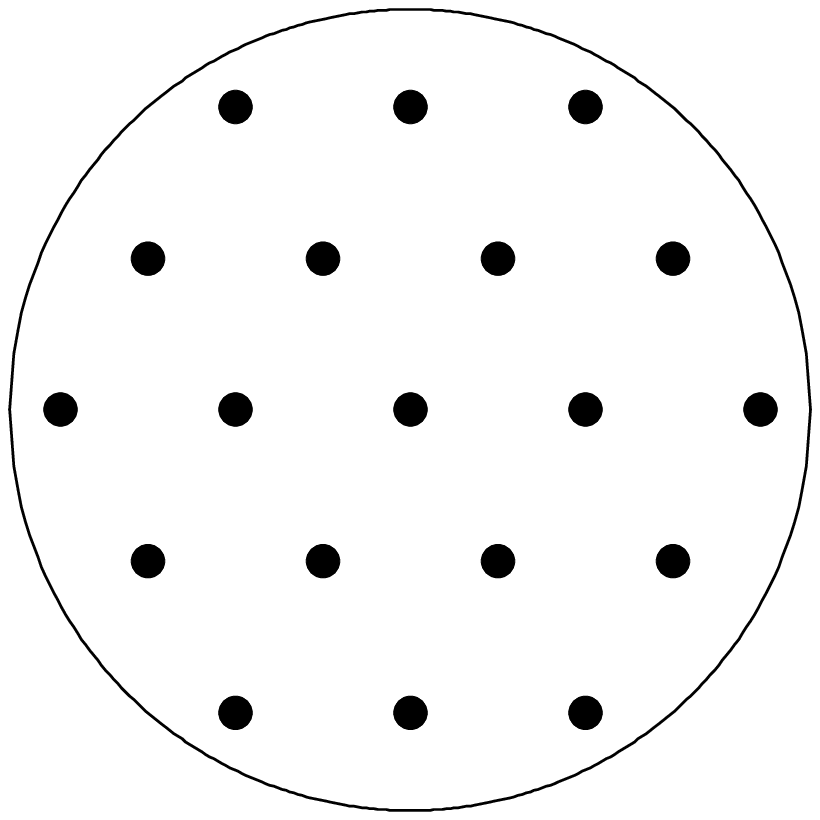} &{\ \ }&
\epsfxsize=3.2cm
\epsffile{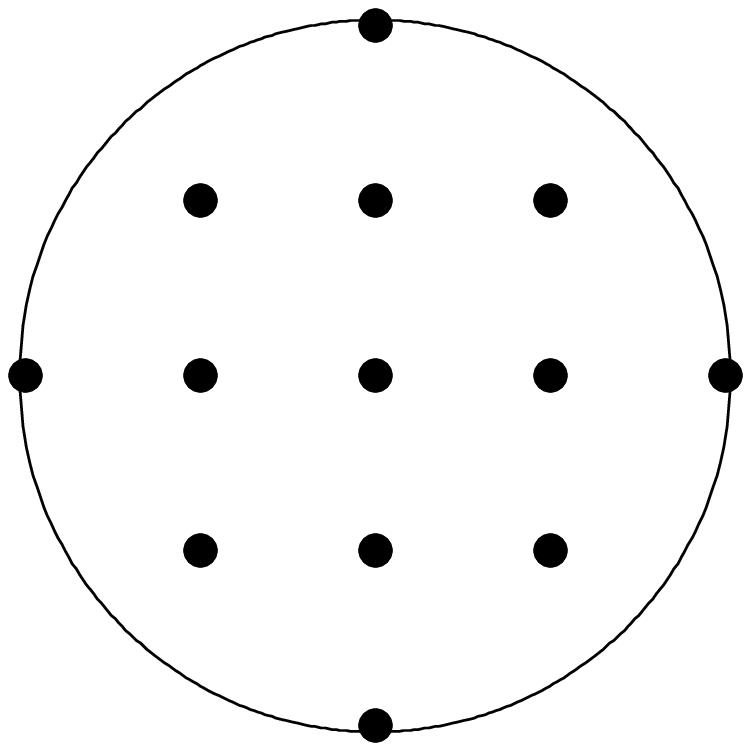} \\
\epsfxsize=3.2cm
\epsffile{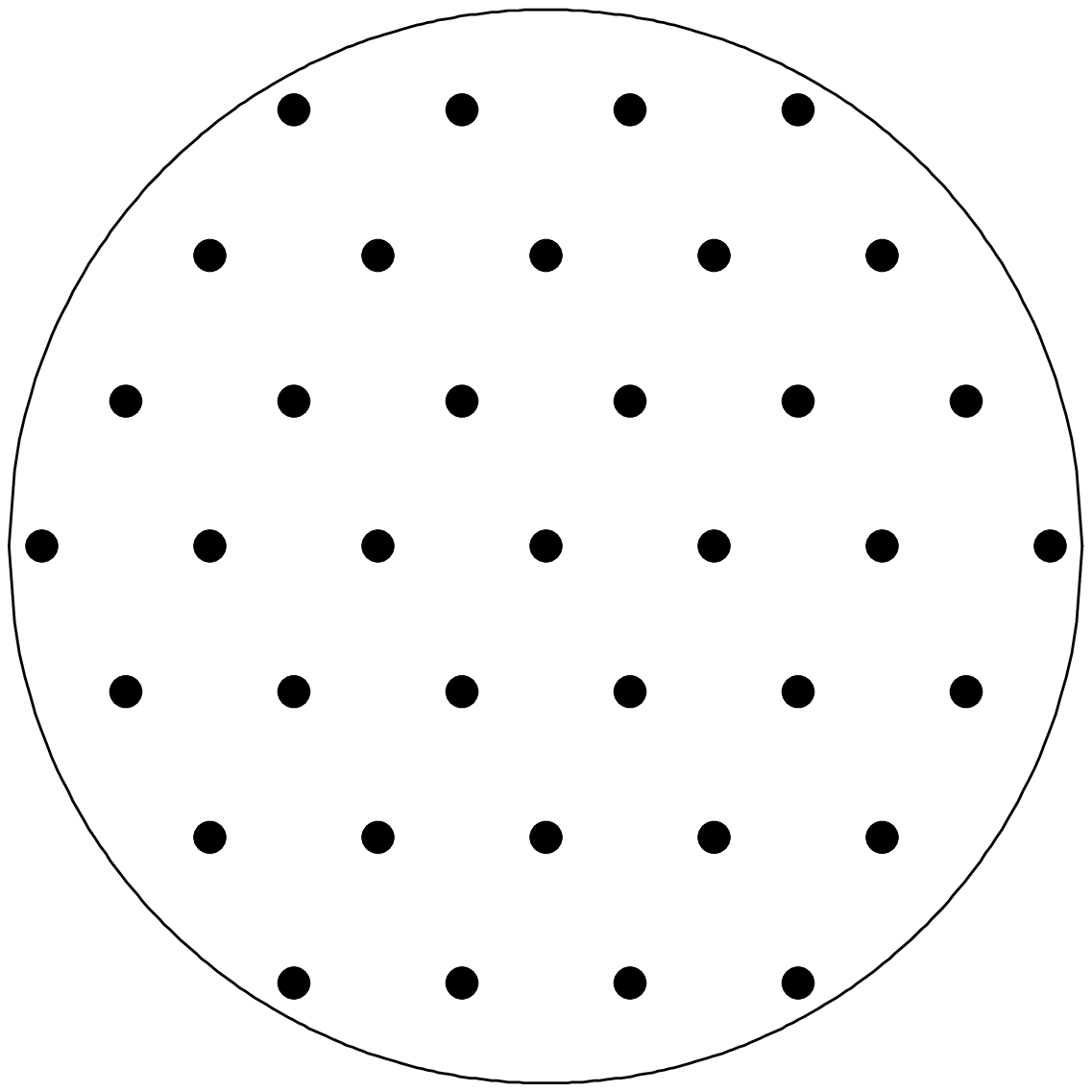} &{\ \ }&
\epsfxsize=3.2cm
\epsffile{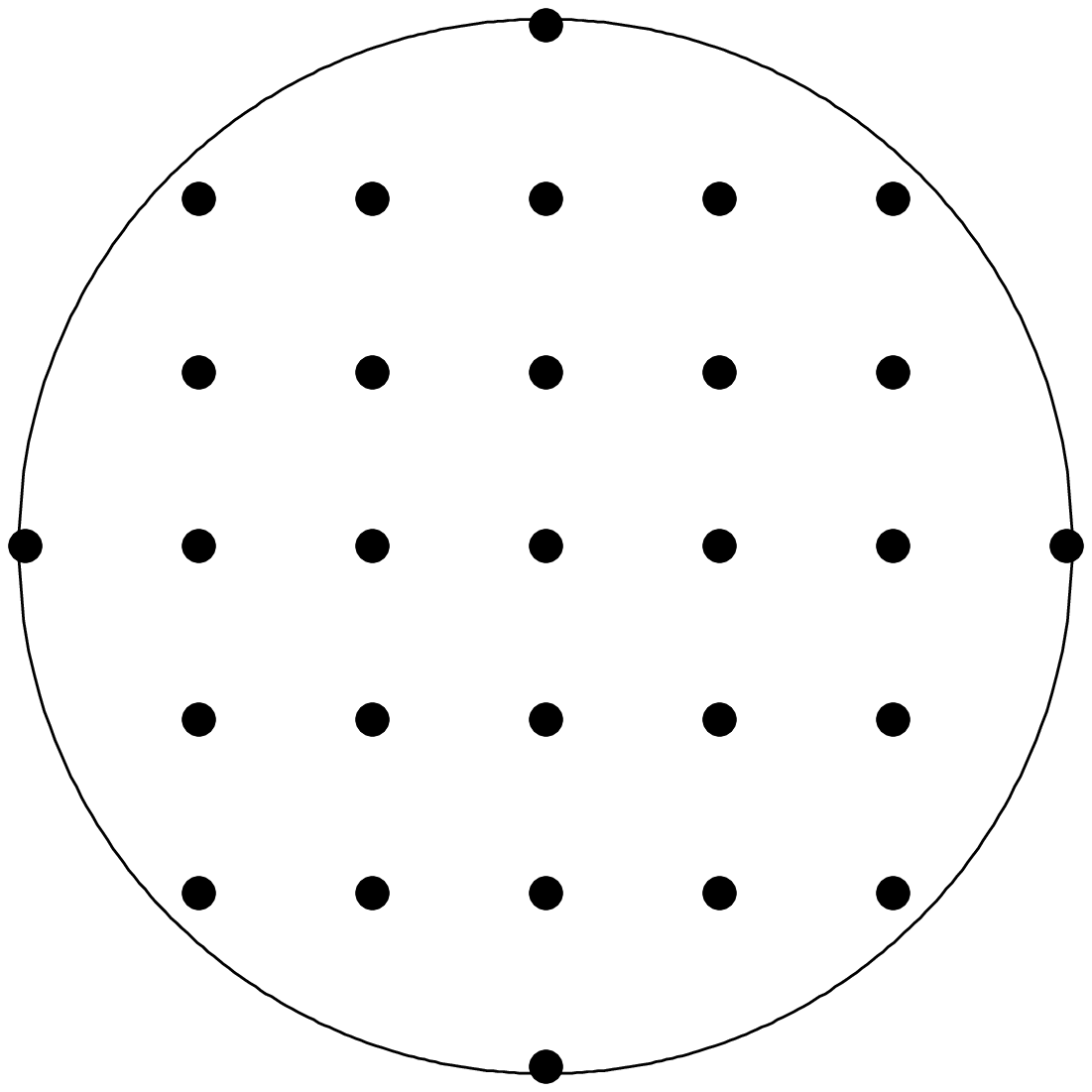} \\
\epsfxsize=3.2cm
\epsffile{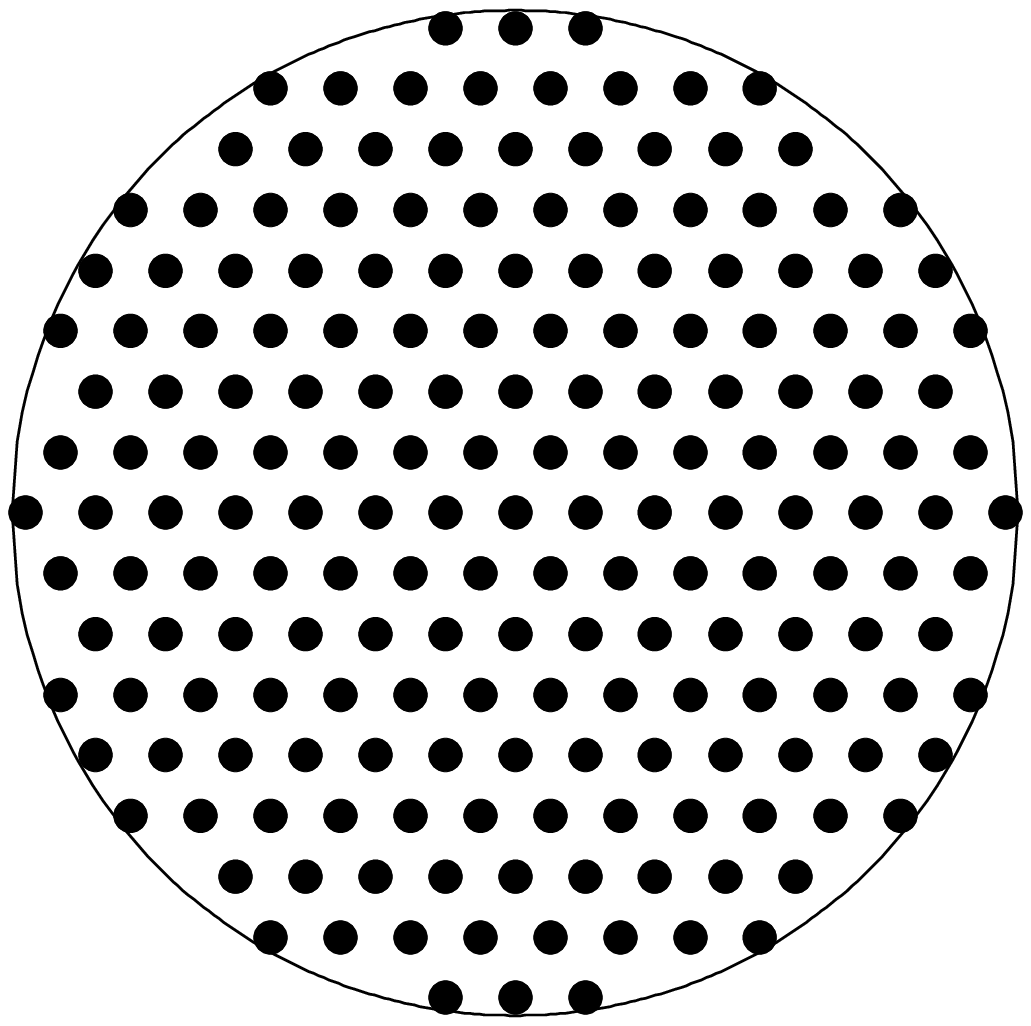} &{\ \ }&
\epsfxsize=3.2cm
\epsffile{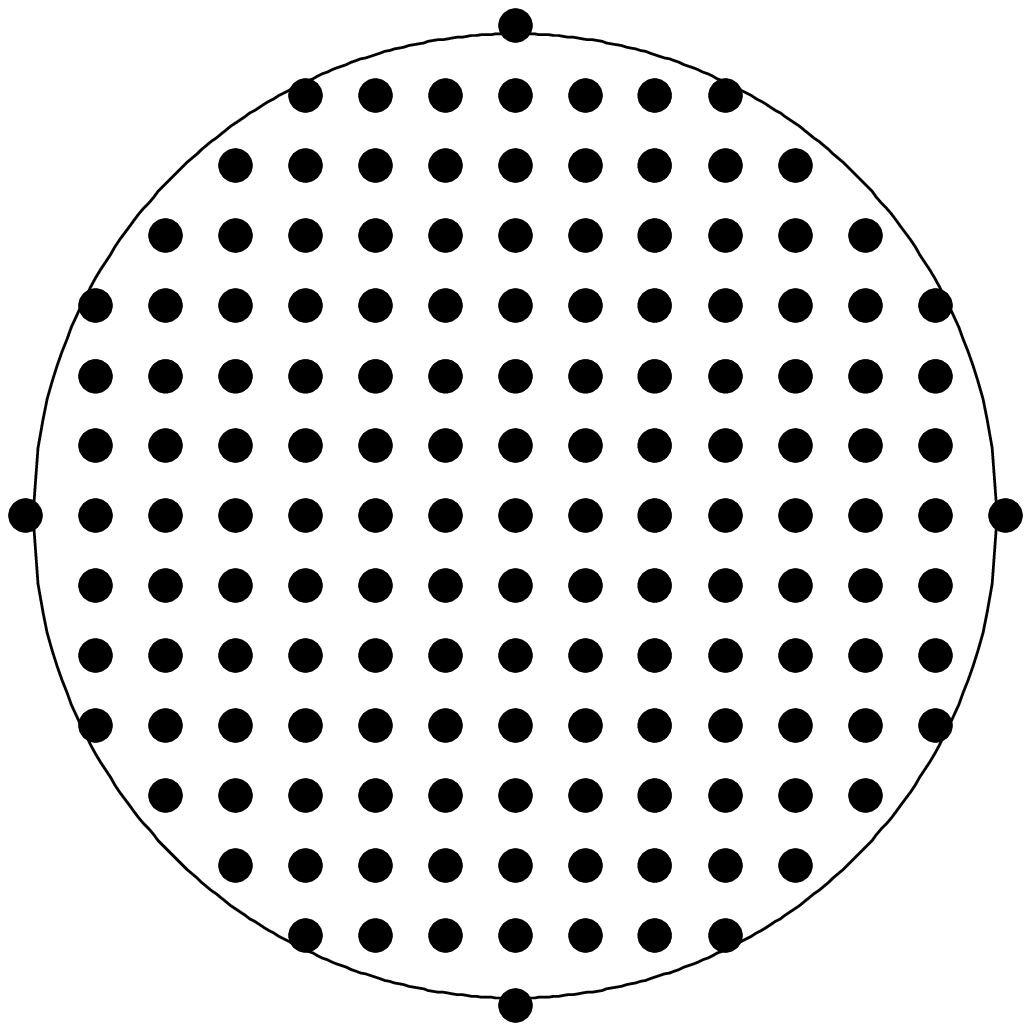} \\
\end{array}
\end{math}
\end{center}
\caption{Examples of finite-size Wigner-lattice configurations which were used in calculations: left plots -- triangular lattice with $N=19$, 37, and 187 lattice points, right plots -- square lattice with $N=13$, 29, and 149 lattice points. Circles show the boundary of the uniform positively charged jellium disks.}
\label{fig-latconfig}
\end{figure}

Here, it is worth discussing the following point. In the Hartree-Fock $N$-electron picture, Eq. (\ref{HF}), electrons occupy the states with $L=0$ to $L_{\max}=N-1$. The last occupied state corresponds to an electron ring with the radius $r_{\max}=\lambda\sqrt{L_{\max}+1}=\lambda\sqrt{N}=R_b\sqrt{\nu}$. At $\nu=1$, the radius $r_{\max}$ exactly coincides with the radius of the positive background $R_b$, and the electron density $n_e({\bf r})$ exponentially decays at $(r-R_b)\gtrsim \lambda$. Sometimes, one draws from here a conclusion, that at $\nu=1$ the states with $L>L_{\max}$ cannot contribute to the expansion (\ref{Nel-func}) (as the inclusion of such states would lead to an essentially non-zero electron density outside the sample), and hence that the Wigner crystal state (\ref{WC}) cannot be considered as a trial wave function at $\nu=1$. This conclusion is however erroneous. The Wigner-crystal wave function (\ref{WC}) is constructed in such a way ($|{\bf R}_j|<R_b$ for all $j$, see Figure \ref{fig-latconfig}), that the electron density outside the sample [at $(r-R_b)\gtrsim \lambda$] is also exponentially small, but in spite of this, the expansion (\ref{WCexpansion}) contains all the states $L_i$, with $0\le L_i <\infty$. 
This remark is valid both for the Wigner crystal state and for all the new proposed states discussed below.

By analogy with the Wigner crystal wave function one can construct an infinite number of trial many-body wave functions which satisfy the electro-neutrality requirement at all magnetic fields and have a clear physical meaning. Consider the function
\be
\Psi^{[N]}({\bf r}_1,{\bf r}_2,\dots,{\bf r}_N)=
\frac{1}{\sqrt{N!}}\det_N
\left |\begin{array}{cccc}
\Phi({\bf r}_1, {\bf R}_1) & 
\Phi({\bf r}_1, {\bf R}_2) &
\dots & 
\Phi({\bf r}_1, {\bf R}_N) \\
\Phi({\bf r}_2, {\bf R}_1) & 
\Phi({\bf r}_2, {\bf R}_2) &
\dots & 
\Phi({\bf r}_2, {\bf R}_N) \\
\dots & \dots & \dots & \dots \\
\Phi({\bf r}_N, {\bf R}_1) & 
\Phi({\bf r}_N, {\bf R}_2) &
\dots & 
\Phi({\bf r}_N, {\bf R}_N)
\end{array}
\right |, 
\label{generalfunction}
\ee
where 
\be
\Phi({\bf r}_i, {\bf R}_j)=\varphi({\bf r}_i-{\bf R}_j) e^{-i\pi{\bf r}_i\cdot({\bf B\times R}_j)/\phi_0}, 
\label{genf}
\ee
and $\varphi({\bf r})$ is an {\it arbitrary} linear combination of the lowest-Landau-level single particle states, 
\be
\varphi({\bf r})=\sum_{L=0}^\infty B_{L} \psi_L({\bf r}).
\label{Phi-functions}
\ee
It is assumed that all the vectors ${\bf R}_j$ in Eq. (\ref{generalfunction}) are different, and the coefficients $B_{L}$ satisfy the normalization condition $\sum_L |B_L|^2 =1$. The function (\ref{generalfunction}) describes the state, in which electrons are ``localized'' near or rotate around the points ${\bf R}_j$ uniformly distributed over the 2D plane with the average density $n_s$. Under this condition the electro-neutrality requirement is fulfilled automatically at all magnetic fields. Note that the vectors ${\bf R}_j$ are free parameters of the theory and are not necessarily given by points of the triangular lattice (\ref{tr-lattice}). They may coincide with points of a square lattice (\ref{sq-lattice}), of any other type of a lattice, or even be randomly distributed over the 2D plane. In general, the configuration of vectors ${\bf R}_j$ should be chosen in view of minimizing the total energy of the system. For example, if effects of disorder are assumed to be important in a given (e.g. low-mobility) sample, the vectors ${\bf R}_j$ may coincide with equilibrium positions of classical point particles randomly distributed over the 2D plane due to the influence of remote donors (the influence of ionized donors on equilibrium positions of point particles in a 2DES was studied in Ref. \cite{Cha94}). Considering different configurations of the vectors ${\bf R}_j$ one can study the interplay between effects of disorder and electron-electron interactions in the integer and the fractional quantum Hall effects. 

The wave functions (\ref{generalfunction}) imply that electrons localized near different ${\bf R}_j$-points are in the same single-particle state (\ref{Phi-functions}). This is also not necessary. One can assume that the single-electron states $\varphi({\bf r}_i-{\bf R}_j)$, localized at different points ${\bf R}_j$, are different. In this case the functions $\Phi({\bf r}_i, {\bf R}_j)$ should be supplied by an additional index $j$. The trial many-body wave functions can then be written in a more general form
\be
\Psi^{[N]}({\bf r}_1,{\bf r}_2,\dots,{\bf r}_N)=
\frac{1}{\sqrt{N!}}\det_N
\left |\begin{array}{cccc}
\Phi_1({\bf r}_1, {\bf R}_1) & 
\Phi_2({\bf r}_1, {\bf R}_2) &
\dots & 
\Phi_N({\bf r}_1, {\bf R}_N) \\
\Phi_1({\bf r}_2, {\bf R}_1) & 
\Phi_2({\bf r}_2, {\bf R}_2) &
\dots & 
\Phi_N({\bf r}_2, {\bf R}_N) \\
\dots & \dots & \dots & \dots \\
\Phi_1({\bf r}_N, {\bf R}_1) & 
\Phi_2({\bf r}_N, {\bf R}_2) &
\dots & 
\Phi_N({\bf r}_N, {\bf R}_N)
\end{array}
\right |,
\label{generalfunction2}
\ee
where $\Phi_j({\bf r}_i, {\bf R}_j)=\varphi_j({\bf r}_i-{\bf R}_j) \exp[-i\pi{\bf r}_i\cdot({\bf B\times R}_j)/\phi_0]$ and 
\be
\varphi_j({\bf r})=\sum_{L=0}^\infty B_{jL} \psi_L({\bf r})
\label{varphiJexpan}
\ee 
(the normalization condition $\sum_L |B_{jL}|^2 =1$ is assumed to be valid for any $j$). If {\it all} the states $\Phi_j$ are different, the vectors ${\bf R}_j$ {\it may be} the same. Like in (\ref{chi0expansion}), the functions $\Phi_j({\bf r}_i, {\bf R}_j)$ can be explicitly expanded in the set of the lowest-Landau-level functions $\psi_L$,

\ba
\Phi_j({\bf r}_i,{\bf R}_j)&=& \sum_{L=0}^\infty B_{jL}\frac{(z_i^*-Z_j^*)^L}{\sqrt{\pi L!}\lambda}e^{-(z_i-Z_j)(z_i^*-Z_j^*)/2-i\pi{\bf r}_i\cdot({\bf B\times R}_j)/\phi_0}\nonumber \\
&=& \sum_{L=0}^\infty B_{jL}\frac{(z_i^*-Z_j^*)^L}{\sqrt{\pi L!}\lambda}e^{-z_iz_i^*/2-Z_jZ_j^*/2 +z_i^* Z_j}=
\sum_{L=0}^\infty D_{jL} (Z_j)\psi_L({\bf r}),
\label{SlatDetExpan}
\ea
where
\be
D_{jL} (Z_j)=\sum_{L^\prime=0}^\infty B_{jL^\prime} \sum_{p=0}^{\min\{L,L^\prime\}}
\frac{(-1)^{L^\prime-p}\sqrt{L!L^\prime!}}{p!(L-p)!(L^\prime-p)!}
Z_j^{L-p}(Z_j^*)^{L^\prime-p} e^{-Z_jZ_j^*/2}.
\ee
Similar to (\ref{WC2expan})--(\ref{WCexpansion}), one can get from here an expansion of the many-body trial wave function (\ref{generalfunction2}) in a set of orthonormal lowest-Landau-level Slater determinants 
(\ref{subset}),
 \be
\Psi^{[N]}({\bf r}_1,{\bf r}_2,\dots,{\bf r}_N)=
 \sum^\infty_{0\le L_1<L_2<\dots <L_N} \det_N
\left |D_{iL_j}(Z_i)\right | \frac 1{\sqrt{N!}}\det_N
\left |\psi_{L_j}({\bf r}_i)
\right |.
\label{generalexpansion}
\ee

The trial many-body wave function (\ref{generalfunction2}) describes a system of particles uniformly distributed over the 2D plane with the average density $n_s$. Each of the particles is in the lowest-Landau-level state. If $\nu>1$, the states with reversed spins, as well as the states from the higher Landau levels may also be included in the expansion (\ref{varphiJexpan}). The vectors ${\bf R}_j$ and the expansion coefficients $B_{jL}$ are free parameters of the theory. They are, in general, some functions of magnetic field, so that when $B$ varies, the system may undergo transitions from, say, a state with the triangular symmetry of the underlying lattice ${\bf R}_j$, to a state with the square symmetry of the lattice, or from a state with one set of coefficients $B_{jL}$, to a state with another set of $B_{jL}$. Some examples of such transitions are discussed in Section \ref{sec:twoparticle}. 

It is also worth noting that the more general form of the trial wave function (\ref{generalfunction}) or (\ref{generalfunction2}) allows one to overcome the problem of negligible overlap of neighbor single-particle states in the Wigner-crystal solution (\ref{WC}) at small $\nu$. As one saw from the discussion in Section \ref{intro}, the underlying triangular symmetry of the Wigner-crystal state {\it could} lead to commensurate energies at certain values of $\nu$, but {\it did not} because of the small overlap of the neighbor gaussians in strong magnetic fields. Using {\it other linear combinations} of the functions $\psi_L$ in Eqs. (\ref{Phi-functions}) or (\ref{varphiJexpan}) (with $B_{jL}\neq\delta_{L,0}$), one can get a large overlap of the neighbors keeping the symmetry of the underlying lattice ${\bf R}_j$. This may lead and does lead to oscillating magnetic-field dependencies of the electron-electron interaction energy of the system, with deep minima in certain intervals of $\nu$.

From this point on I will consider a few examples of possible trial wave functions, assuming that (i) effects of disorder are negligible, so that the vectors ${\bf R}_j$ form a lattice; (ii) the lattice of the vectors ${\bf R}_j$ has either the triangular (\ref{tr-lattice}) or the square (\ref{sq-lattice}) symmetry; (iii) the coefficients $B_{jL}$ in the expansion (\ref{varphiJexpan}) do not depend on $j$, i.e. all single-particle states centered at different lattice points are the same, and (iv) the expansion (\ref{varphiJexpan}) consists of the only one term, $B_{jL^\prime}=\delta_{L,L^\prime}$. Thus, the trial many-body wave functions to be considered in this paper are 

\be
\Psi_{L}^{[N]}=\frac 1{\sqrt{N!}}\det_N|\chi_{L}({\bf r}_i,{\bf R}_j)|,
\label{PsiL}
\ee
where the functions $\chi_L$ are defined by Eq. (\ref{chi-functions}). The functions $\Psi^{[N]}_L$ are supplied by an additional subscript $L$ which labels different variational solutions of the problem. They describe a system of electrons, localized at ($L=0$) or rotating around ($L>0$) the points ${\bf R}_j$ of the triangular or the square lattice. The state $\Psi_{L=0}^{[N]}$ is the Wigner crystal state \cite{Maki83}. The functions $\chi_{L}({\bf r}_i,{\bf R}_j)$ with different ${\bf R}_j$ are not orthogonal to each other, so that the trial functions $\Psi_{L}^{[N]}$ take into account electron-electron correlations in an approximation far beyond the Hartree-Fock one.

It would certainly be meaningless to discuss the complicated many-body wave functions in the form (\ref{PsiL}) or (\ref{generalfunction2}), if there were no possibility to calculate the energy or other physical properties of these states. Fortunately, there exists {\it an exact analytical method} of calculating expectation values of {\it practically any physical quantity}, including the density, pair-correlation function, energy, et cetera, in the states (\ref{PsiL}), (\ref{generalfunction}), or (\ref{generalfunction2}). I present this method in the next Section.

\section{properties of the many-body states}
\label{sec:properties}

In order to calculate expectation values of different physical quantities in the states $\Psi_L^{[N]}$ one needs to know the norm $\langle\Psi_L^{[N]}|\Psi_L^{[N]}\rangle$, and the matrix elements of arbitrary single-particle and two-particle operators. This can be done with the help of general results formulated in Appendix \ref{app:manybody}. 

All the specific calculations below have been performed in a finite-size Wigner-cluster configuration, with the radius of the positively charged disk $R_b$, determined by Eq. (\ref{radiusR}), and $N$ lattice points (\ref{tr-lattice}) or (\ref{sq-lattice}), placed within a circle of the radius $R_b$, Figure \ref{fig-latconfig}. Configurations with $N=7$, 19, 37, $\dots$ and $N=5$, 13, 29, $\dots$ lattice points have been considered in the triangular and the square lattice, respectively. 

\subsection{The norm}
\label{sec:norm}

According to (\ref{S-func}), the norm $\langle\Psi_L^{[N]}|\Psi_L^{[N]}\rangle$ is determined by the formula

\be
\langle\Psi_L^{[N]}|\Psi_L^{[N]}\rangle=\det_N \left | {\cal S}_N^{LL} \right |,
\label{norma}
\ee
where ${\cal S}_N^{LL}$ is the $N\times N$ matrix of overlap integrals with the elements
\be
S_{ij}^{LL}\equiv \langle \chi^L_{ai}|\chi^L_{aj}\rangle =\exp\left[-\frac{\xi_i\xi_i^*+\xi_j\xi_j^*}2+\xi_i^*\xi_j\right]
L_L\left(\eta_{ij}\eta_{ij}^*\right).
\label{SijLL}
\ee
Here $L_L$ are the Laguerre polynomials, the complex variables $\xi_i$ are related to vectors ${\bf R}_i=(X_i,Y_i)$,
\be
\xi_i=(X_i+iY_i)/\lambda,
\label{xi-define}
\ee
and $\eta_{ij}=\xi_i-\xi_j$.

\begin{figure}
\begin{center}
\begin{math}
\epsfxsize=8.5cm
\epsffile{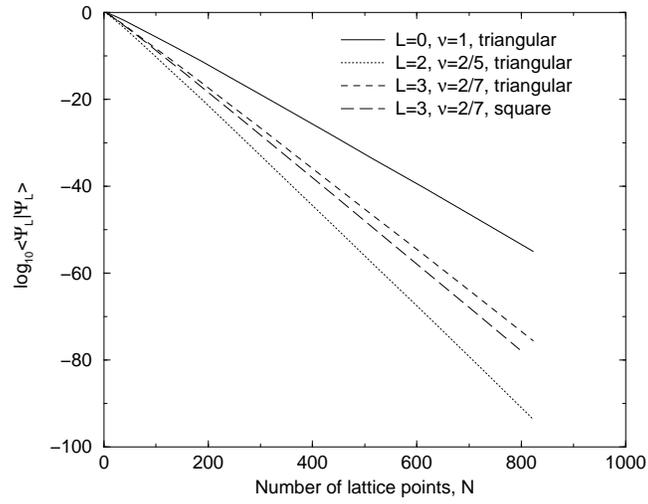}
\end{math}
\end{center}
\caption{Logarithm of the norm $\log \langle\Psi_L|\Psi_L\rangle$ as a function of the number of lattice points involved in calculations, for a few different $L$ and $\nu$, for a triangular and a square symmetry of the lattice. }
\label{fig:norm}
\end{figure}

The norm $\langle\Psi_L^{[N]}|\Psi_L^{[N]}\rangle$ exponentially tends to zero with growing $N$,
\be
\langle\Psi_L^{[N]}|\Psi_L^{[N]}\rangle\sim e^{-\alpha N} {\ \ \rm at\ } N\to\infty,
\label{normtozero}
\ee
where $\alpha$ depends on the overlap integrals, see Figure \ref{fig:norm}. This property does not lead to any problem in calculating the expectation value of a physical quantity $O$, as in the formula $O=\langle\Psi_L^{[N]}|\hat O|\Psi_L^{[N]}\rangle/\langle\Psi_L^{[N]}|\Psi_L^{[N]}\rangle$ both the nominator and the denominator contain similar integrals, both tend to zero in the limit $N\to\infty$, while the ratio $O$ tends to a constant. 

\subsection{Single-particle operators}
\label{sec:singleparticle}

Matrix elements of single-particle operators 
\be
\hat O_1=\sum_{i=1}^N \hat o_1({\bf r}_i)
\label{O1}
\ee
can be found with the help of Eqs. (\ref{Hone}) and (\ref{HoneSP}). In order to calculate them one needs to know the matrix ${\cal S}_N^{LL}$ (Section \ref{sec:norm}) and single-particle matrix elements $\langle \chi^L_{ai}|\hat o_1({\bf r}_a)|\chi^L_{aj}\rangle$ of a specific operator $\hat o_1$. Due to a simple analytical form of the functions $\chi_L$, the single-particle matrix elements $\langle \chi^L_{ai}|\hat o_1({\bf r}_a)|\chi^L_{aj}\rangle$ can be analytically calculated for practically any physical quantity of interest. A few examples are presented below.

Let $\hat o_1$ be the kinetic energy operator. For the single-particle  matrix element I get

\be
\langle \chi^L_{ai}|\hat k_a|\chi^L_{aj}\rangle=\frac{1}{2m^\star}\int d{\bf r}\chi^*_L({\bf r},{\bf R}_i)
\left(\hat {\bf p}+\frac {|e|}{2c}{\bf B}\times{\bf
r}\right)^2
\chi_L({\bf r},{\bf R}_j)=\frac{\hbar\omega_c}2S_{ij}^{LL},
\ee
so that immediately

\be
\frac{\langle \Psi^{[N]}_L|\hat K|\Psi^{[N]}_L\rangle }{\langle \Psi^{[N]}_L|\Psi^{[N]}_L\rangle }=N\frac{\hbar\omega_c}2,
\ee
-- the kinetic energy per particle in the state $\Psi_L^{[N]}$ is $\hbar\omega_c/2$. 

A less trivial example of a single-particle operator is the electron density

\be
\hat n({\bf r})=\sum_i \delta({\bf r-r}_i)=\sum_{\bf G}\hat n_{\bf G}e^{i{\bf G\cdot r}},
\ee 
${\bf G}$ are reciprocal lattice vectors. It can be calculated either directly or via its Fourier transform. In the former case one needs the matrix elements

\be
n_{ij}^{LL}({\bf r})=\langle \chi^L_{ai}|\delta({\bf r-r}_a)|\chi^L_{aj}\rangle
=\chi^*({\bf r},{\bf R}_i)\chi({\bf r},{\bf R}_j),
\label{Nmelement}
\ee
in the latter case -- Fourier components

\ba
S_{ij}^{LL}({\bf q})&\equiv&\langle \chi^L_{ai}|e^{i{\bf q\cdot r}_a}|\chi^L_{aj}\rangle \nonumber \\
&=&\exp\left[-\frac{\xi_i\xi_i^*+\xi_j\xi_j^*}2+(\xi_i^*+i\gamma^*/2)(\xi_j+i\gamma/2)\right]
L_L\left[(\eta_{ij}-i\gamma/2)(\eta_{ij}^*+i\gamma^*/2)\right],
\label{SfourierLL}
\ea
where $\gamma=(q_x+iq_y)\lambda$. Together with (\ref{Hone}) and (\ref{HoneSP}), Eqs. (\ref{Nmelement}) and (\ref{SfourierLL}) give exact analytic expressions for the electron density and its Fourier components in the states $\Psi_L^{[N]}$.

\begin{figure}
\begin{center}
\begin{math}
\begin{array}{cc}
\epsfxsize=7.0cm
\epsffile{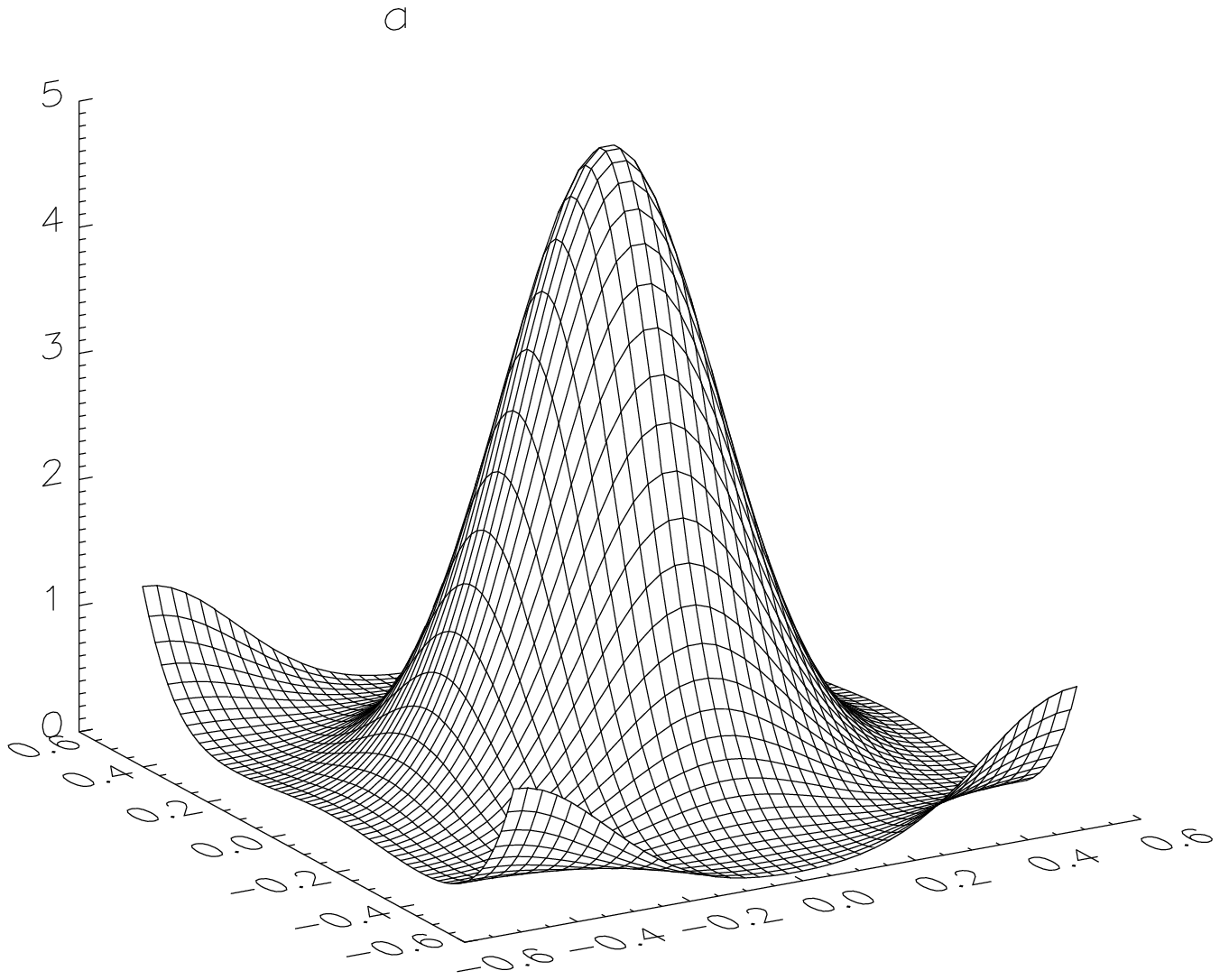} &
\epsfxsize=7.0cm
\epsffile{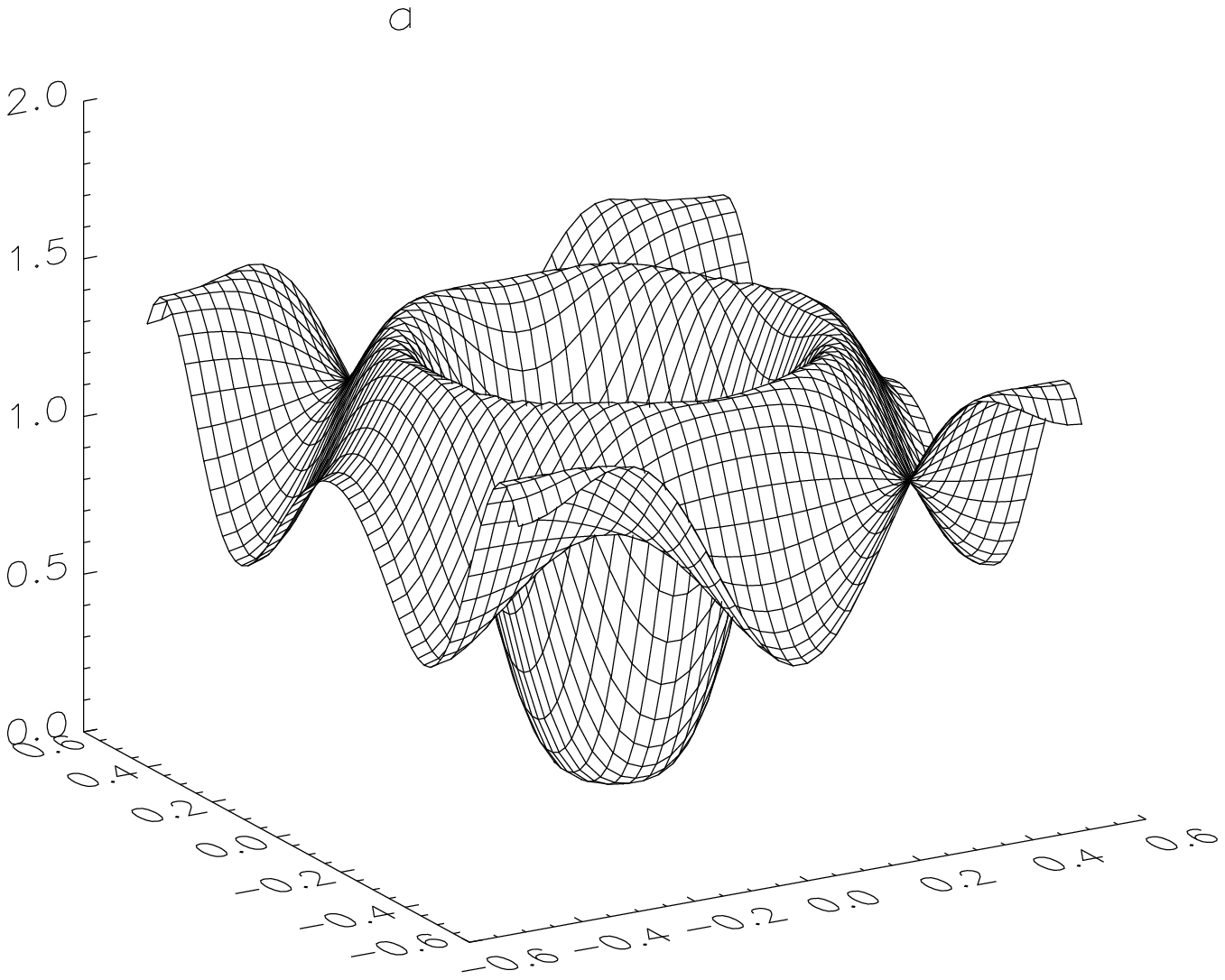} \\
\epsfxsize=7.0cm
\epsffile{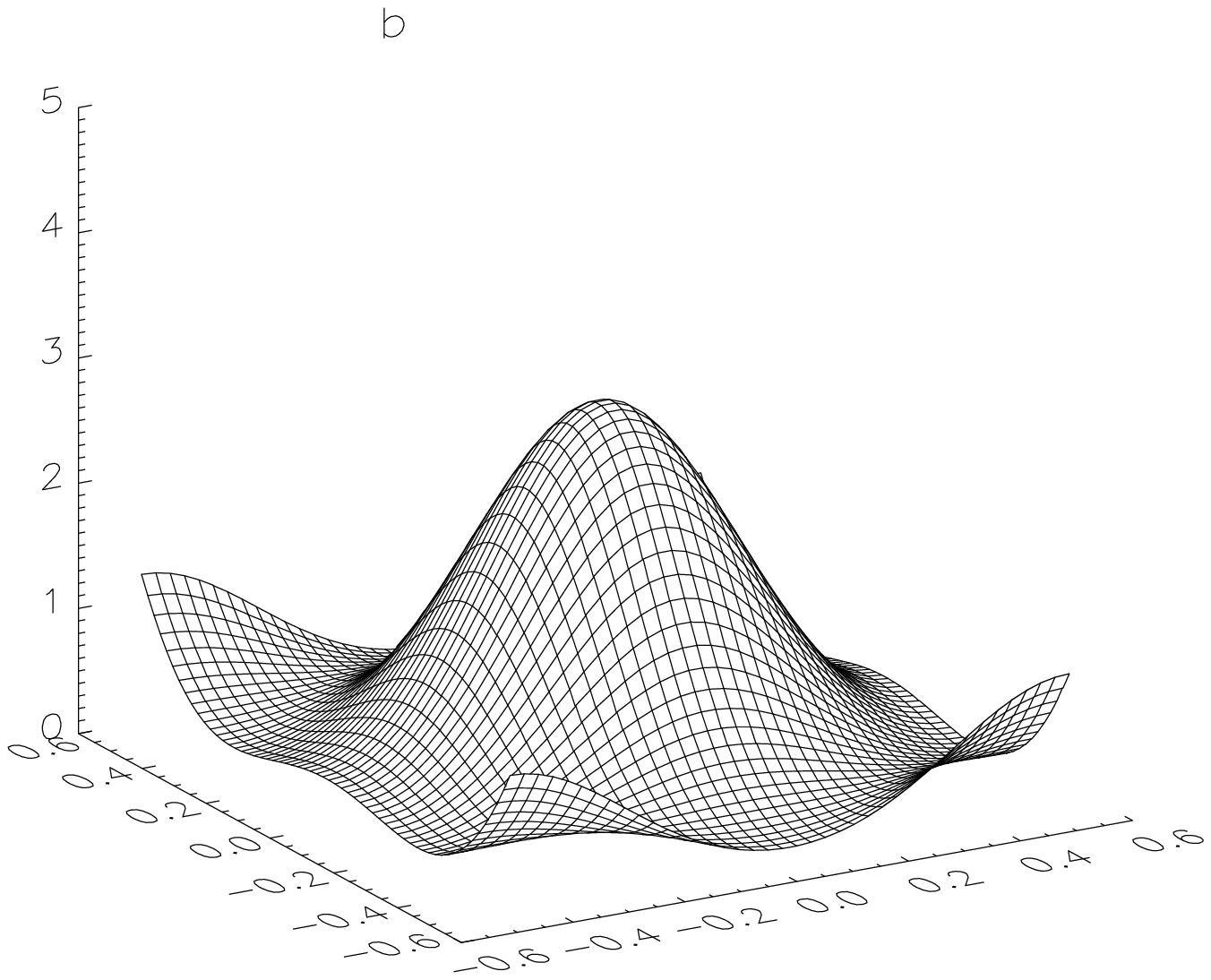} &
\epsfxsize=7.0cm
\epsffile{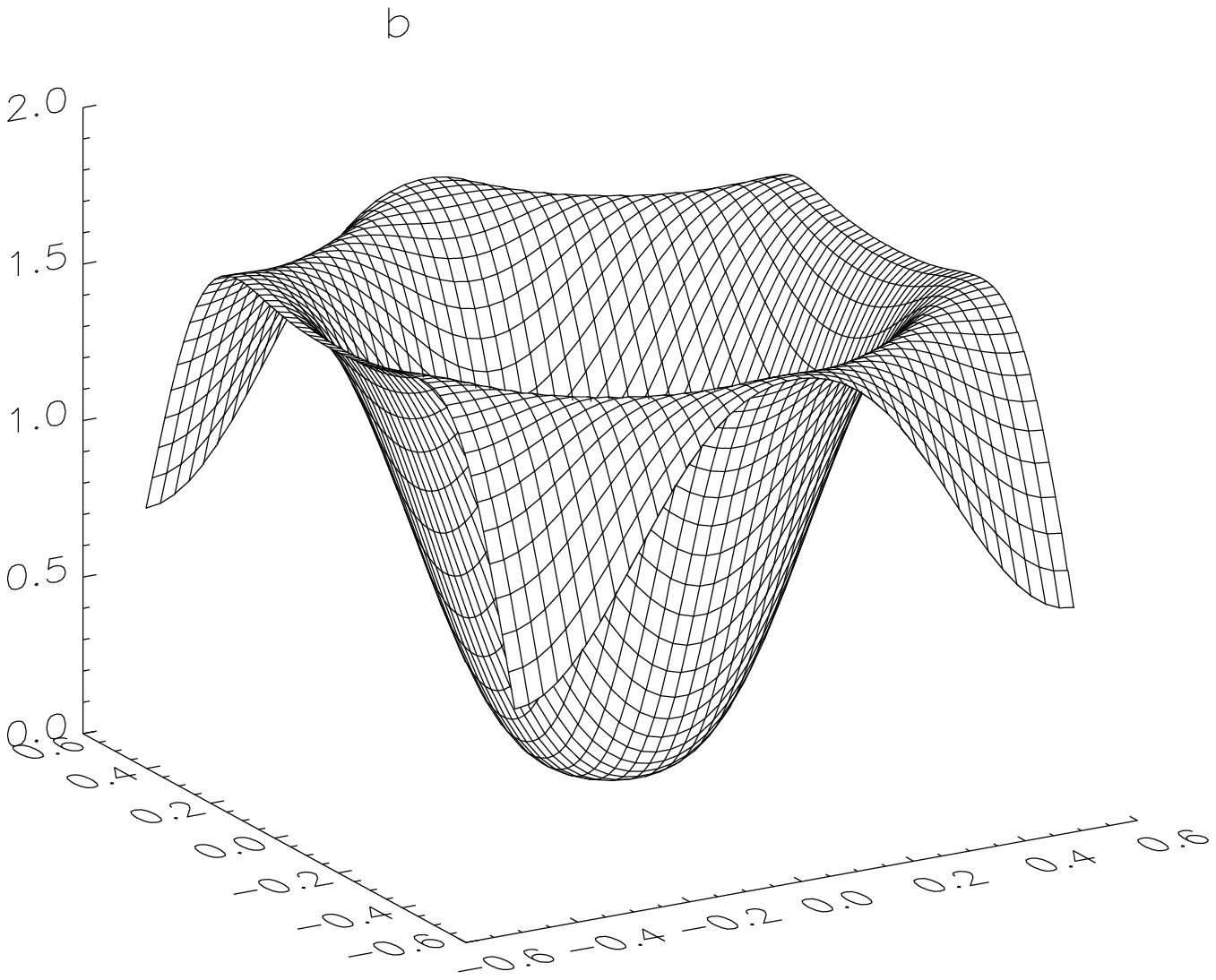} \\
\epsfxsize=7.0cm
\epsffile{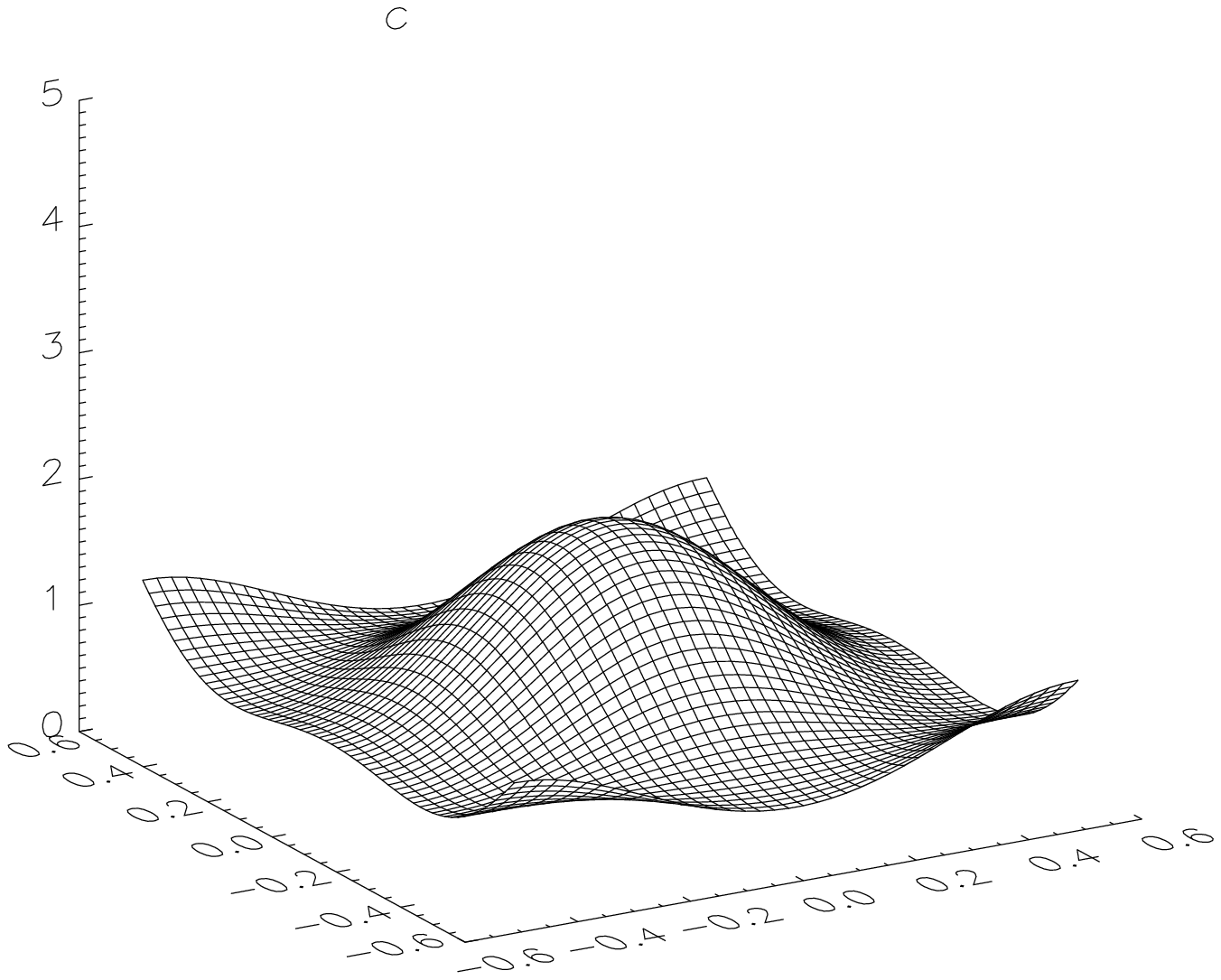} &
\epsfxsize=7.0cm
\epsffile{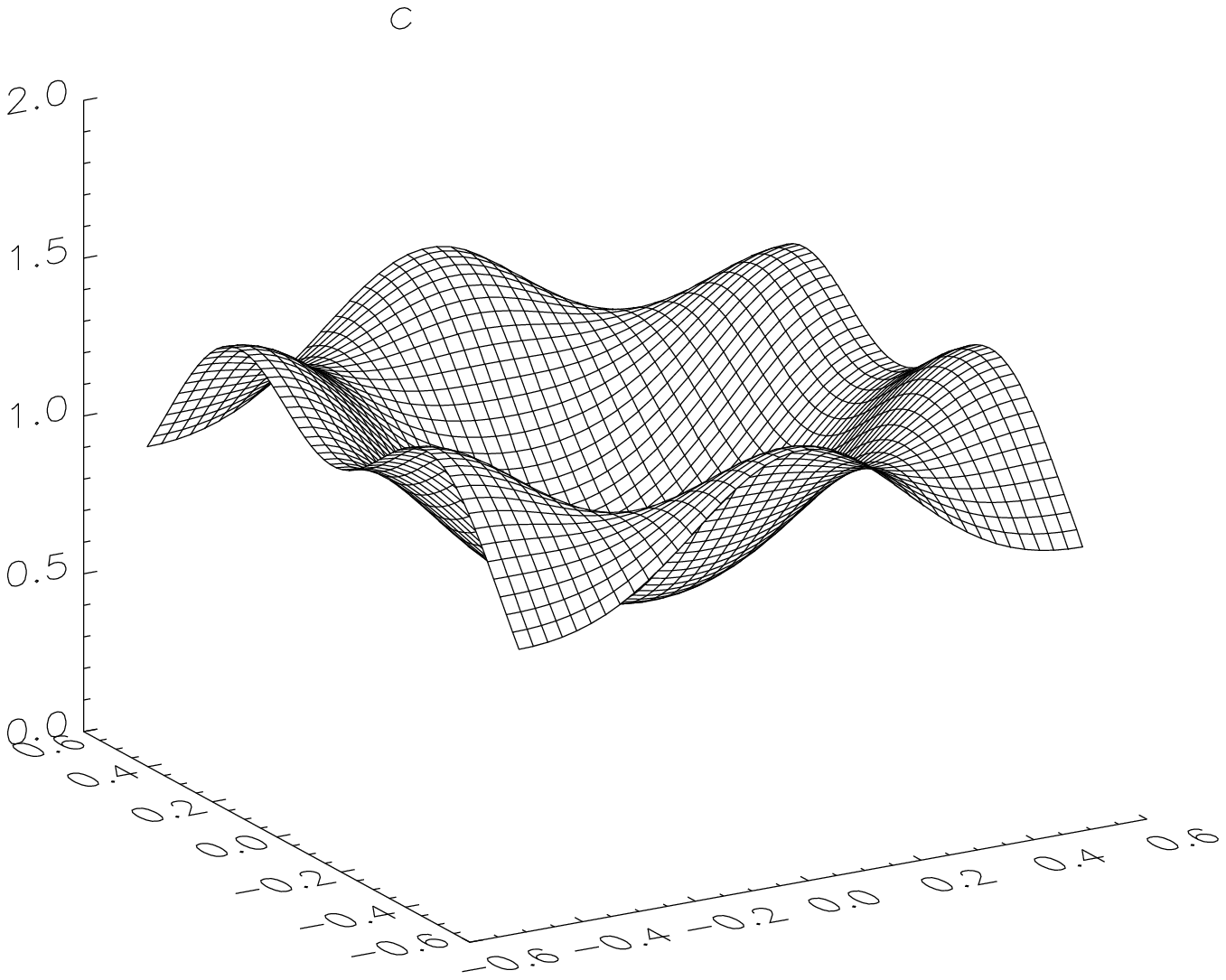} \\
\epsfxsize=7.0cm
\epsffile{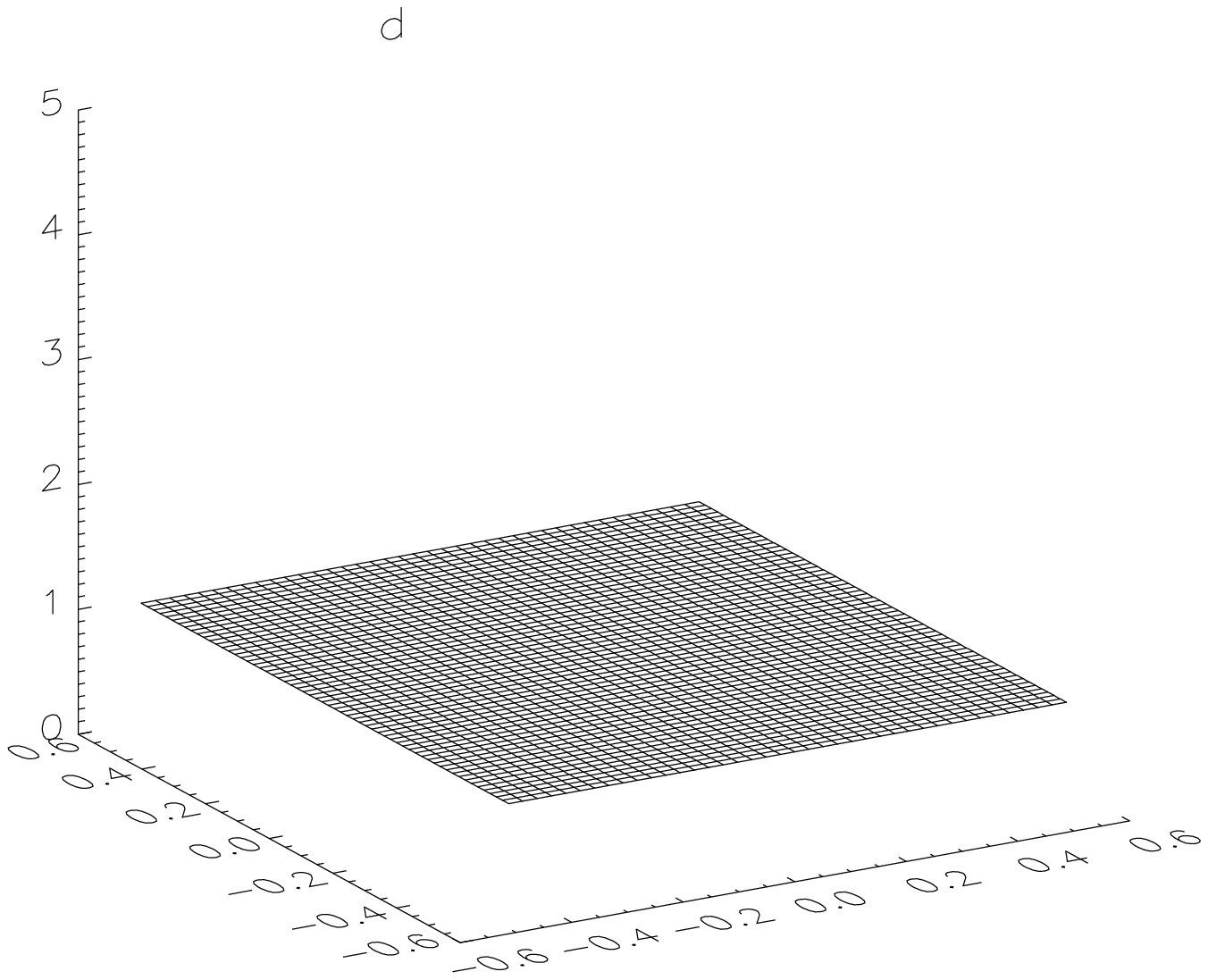} &
\epsfxsize=7.0cm
\epsffile{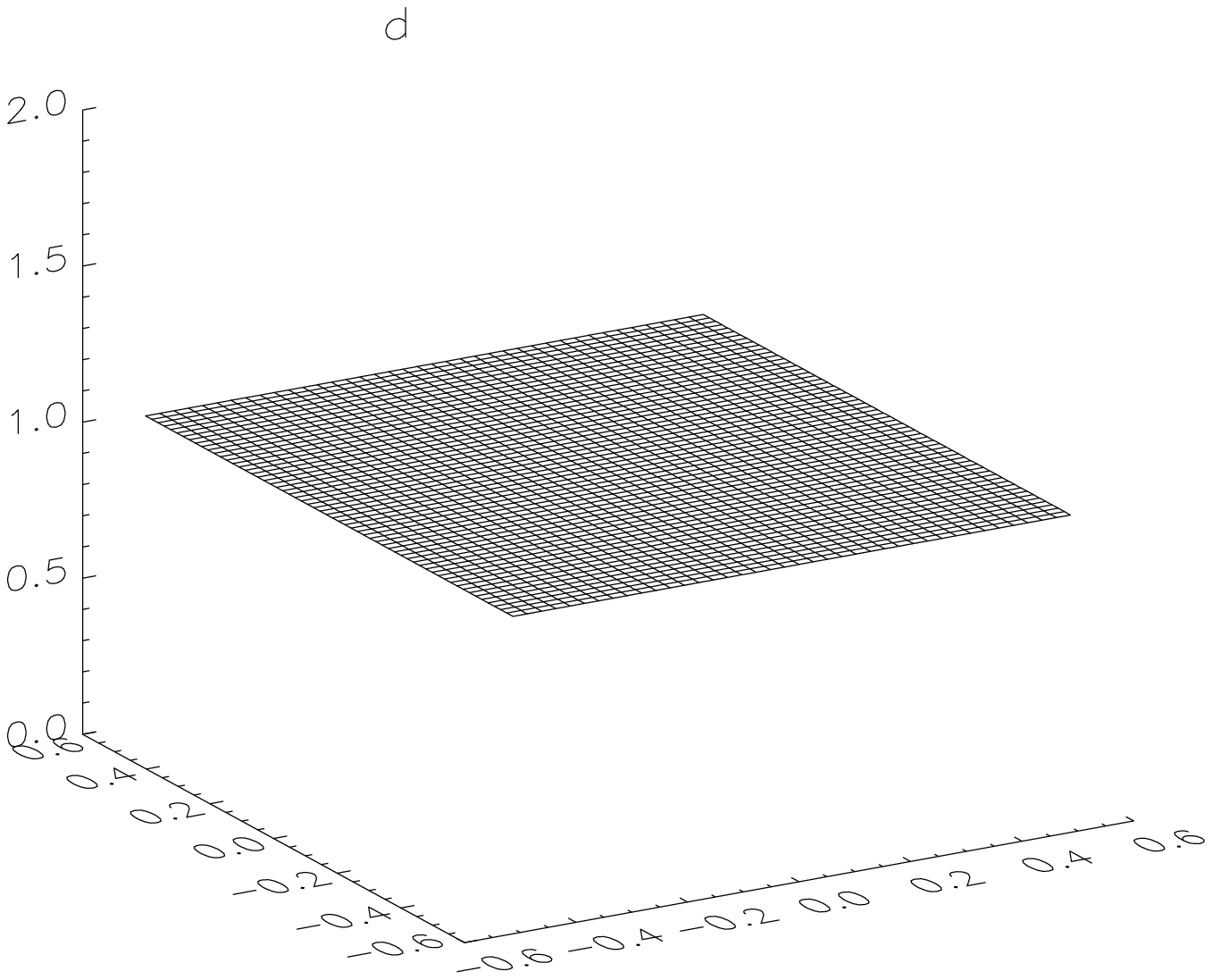} \\
\end{array}
\end{math}
\end{center}
\caption{Calculated normalized electron density $n_e({\bf r})/n_s$ in the states $L=0$ (left panels) and $L=2$ (right panels), with a triangular lattice of vectors ${\bf R}_j$, at the Landau level filling factors (a) $\nu=1/5$, (b) $\nu=1/3$, (c) $\nu=1/2$, and (d) $\nu=1$.}
\label{fig-density}
\end{figure}

\begin{figure}
\begin{center}
\begin{math}
\epsfxsize=8.5cm
\epsffile{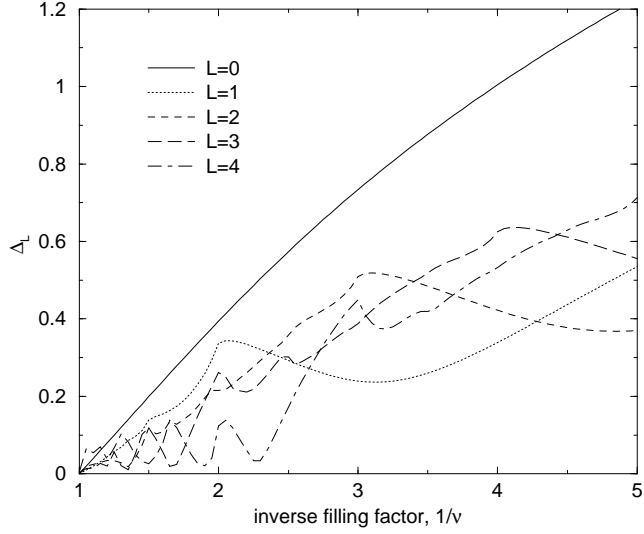}
\end{math}
\end{center}
\caption{Square-root deviation $\Delta_L$ of the electron density from the average value $n_s$, Eq. (\ref{sqrdev}), as a function of the inverse Landau-level filling factor $\Phi=\nu^{-1}=B/n_s\phi_0$, for a few lowest $\Psi_L$-states in a system with a triangular lattice of points ${\bf R}_j$. The results are obtained by extrapolation to the thermodynamic limit $N\to\infty$ as described in the text.}
\label{fig:sqr-deviation}
\end{figure}

Figure \ref{fig-density} exhibits coordinate dependencies of the normalized electron density, $n_e({\bf r})/n_s$, in the states $\Psi_{L=0}$ and $\Psi_{L=2}$, with a triangular lattice of vectors ${\bf R}_j$, at a few values of the Landau level filling factor $\nu$. The Fourier components of the electron density have been calculated for a large number of configurations (Figure \ref{fig-latconfig}, left panel) with $N$ up to $\sim 1000$ (they are with a good accuracy proportional to $1/\sqrt{N}$), and then extrapolated to $N\to\infty$. Figure \ref{fig-density} thus shows the electron density in the thermodynamic limit. An interesting finding is that, in spite of the underlying periodic symmetry of the lattice  ${\bf R}_j$, the density of electrons $n_e({\bf r})=1/\pi\lambda^2$ is uniform in the limit $\nu=1$ of a completely filled lowest Landau level. This is due to a strong overlap of the neighbor single-particle states at $\nu=1$, and seems to be valid at all $L$, at least for a triangular lattice of points ${\bf R}_j$. I have calculated the square-root deviation of the electron density from the average value,

\be
\Delta_L= \left( \sum_{\bf G\neq 0} |n_{\bf G}^{LL}/n_s|^2 \right)^{1/2},
\label{sqrdev}
\ee
as a function of magnetic field $B$ for a few lowest $L$-states in the system with a triangular lattice of points ${\bf R}_j$, Figure \ref{fig:sqr-deviation}. The functions $\Delta_{L>0}(B)$ have a complicated oscillating dependence on the magnetic field, due to the overlap of electron rings rotating around the lattice points, and all the curves $\Delta_L(B)$ tend to zero at $\nu\to 1$. 

Similarly, one can calculate expectation values of other single-particle operators  in the states $L$ at any magnetic field. Results of a more complete analysis of the single-particle properties of the states $\Psi_L$ will be reported in subsequent publications.

\subsection{Two-particle operators}
\label{sec:twoparticle}

Matrix elements of two-particle operators 
\be
\hat O_2=
\sum_{i=1}^N \sum_{j=1 (j\neq i)}^N \hat 
o_2({\bf r}_i,{\bf r}_j)
\label{O2}
\ee
can be calculated with the help of Eqs. (\ref{Htwo}) and (\ref{2partme}). In order to calculate them one needs to know, apart from the matrix ${\cal S}_N^{LL}$, the two-particle matrix elements (\ref{2partme}) of a specific operator $\hat o_2$. For instance, the matrix elements of the pair-correlation function can be easily calculated. I will focus below on the most important two-particle operator in the problem -- the electron-electron interaction energy $V_{ee}$, Eq. (\ref{Vee}).

According to Eq. (\ref{Htwo}) one gets

\be
\langle\Psi_L^{[N]}|V_{ee}|\Psi^{[N]}_{L}\rangle=\frac 12
\sum_{i=1}^N\sum_{j=1}^N
\sum_{k=1}^N\sum_{l=1}^N(-1)^{i+j+k+l}
{\rm sgn}(i-k){\rm sgn}(j-l)
V_{ijkl}^{LL}\det_{N-2} | {\cal S}_{N-2}^{LL}|^{row\neq i,k}_{column\neq j,l},
\label{Veeme}
\ee
where the $(N-2)\times (N-2)$ matrix ${\cal S}_{N-2}^{LL}|^{row\neq i,k}_{column\neq j,l}$ is formed from the matrix ${\cal S}_N^{LL}$, Eq. (\ref{SijLL}), by deleting the rows $i,k$ and the columns $j,l$. The matrix elements $V_{ijkl}^{LL}$ are determined by

\be
V_{ijkl}^{LL}=\int\int d{\bf r}_ad{\bf r}_b (\chi_{ai}^L)^*\chi_{aj}^{L}\frac{e^2}{|{\bf r}_a-{\bf r}_b|}(\chi_{bk}^L)^*\chi_{bl}^{L},
\label{Vijkl_LL}
\ee
have the following symmetry properties
\be
V_{ijkl}^{LL}=V_{klij}^{LL},{\ \ \ \ }V_{ijkl}^{LL}=(V_{jilk}^{LL})^*,
\label{Vproperties}
\ee
and are calculated in Appendix \ref{appVijkl}. With the help of (\ref{Vproperties}), Eq. (\ref{Veeme}) can be presented in the form

\ba
&&\langle\Psi_L^{[N]}|V_{ee}|\Psi_{L}^{[N]}\rangle=\frac 12
\sum_{i=1}^N\sum_{j=1(\neq i)}^N
(V_{iijj}^{LL}-
V_{ijji}^{LL})\det_{N-2} |{\cal S}_{N-2}^{LL}|^{row\neq i,j}_{column\neq i,j}\nonumber \\
&+&
\sum_{i=1}^N
\sum_{j=1(\neq i)}^N\sum_{k=1(\neq i,j)}^N(-1)^{j+k}
{\rm sgn}(i-j){\rm sgn}(i-k){\rm Re}\left[
(V_{iijk}^{LL}-V_{ikji}^{LL})\det_{N-2} |
{\cal S}_{N-2}^{LL}|^{row\neq i,j}_{column\neq i,k}\right]\nonumber \\
&+&\frac 12
\sum_{i=1}^N\sum_{j=1 (\neq i)}^N
\sum_{k=1(\neq i,j)}^N\sum_{l=1(\neq i,j,k)}^N(-1)^{i+j+k+l}
{\rm sgn}(i-k){\rm sgn}(j-l) 
{\rm Re}\left[
V_{ijkl}^{LL}\det_{N-2} |
{\cal S}_{N-2}^{LL}|^{row\neq i,k}_{column\neq j,l}\right],
\label{three-contrib}
\ea
where the first, second and the third lines correspond to the two-site (direct and exchange), three-site and four-site contributions to the total energy of the system, respectively. Dividing this result by the norm (\ref{norma}), I get the expectation value of the energy of electron-electron interaction in the state $\Psi_L^{[N]}$, 

\ba
\langle V_{ee}\rangle_L^{[N]}&\equiv& \frac{\langle\Psi_L^{[N]}|V_{ee}|\Psi_{L}^{[N]}\rangle}{\langle\Psi_L^{[N]}|\Psi_{L}^{[N]}\rangle}=
\underbrace{\frac 1{2}
\sum_{i=1}^N\sum_{j=1(\neq i)}^N
V_{iijj}^{LL}}_{Hartree\ term}
-\frac 1{2}
\sum_{i=1}^N\sum_{j=1(\neq i)}^N
\left[V_{iijj}^{LL}\left(1-\alpha^{ij}_{ij}\right)+
V_{ijji}^{LL}\alpha^{ij}_{ij}\right]\nonumber \\
&+&
\frac 1{2} \sum_{i=1}^N
\sum_{j=1(\neq i)}^N\sum_{k=1(\neq i,j)}^N(-1)^{j+k}
{\rm sgn}(i-j){\rm sgn}(i-k){\rm Re}\left[2
(V_{iijk}^{LL}-V_{ikji}^{LL})\alpha^{ij}_{ik}\right]\nonumber \\
&+&\frac 1{2}
\sum_{i=1}^N\sum_{j=1(\neq i)}^N
\sum_{k=1(\neq i,j)}^N\sum_{l=1(\neq i,j,k)}^N(-1)^{i+j+k+l}
{\rm sgn}(i-k){\rm sgn}(j-l){\rm Re}\left[
V_{ijkl}^{LL}\alpha^{ik}_{jl}\right],
\label{EEpp}
\ea
where the first (underbraced) term is the Hartree energy, see Appendix \ref{app:Hartree}, and the rest -- the exchange-correlation contribution. The factors $\alpha^{ik}_{jl}$ in (\ref{EEpp}) are determined by 

\be
\alpha^{ik}_{jl}=\frac { 
\det_{N-2} |{\cal S}_{N-2}^{LL}|^{row\neq i,k}_{column\neq j,l}}{\det_N |{\cal S}_N^{LL}|}.
\label{alphafactor}
\ee

Electron-electron interaction energy $\langle V_{ee}\rangle_L^{[N]}$ grows as $N^{3/2}$ at $N\to\infty$ (the energy of a charged disk). In order to compensate this growth, I subtract the energy of the positive background 

\be
V_{bb}=\frac{e^2\sqrt{n_s}}\kappa \frac {8N}3\sqrt{\frac N\pi},
\ee
and divide the result by $N$ (to get the energy per particle). Thus calculated estimate of the electron-electron interaction energy 

\be
\epsilon_L^{ee}(N)=\frac{\langle V_{ee}-V_{bb}\rangle_L^{[N]}}N
\label{eee}
\ee
differs from the cohesive energy of the system by the term $\langle V_{eb}+2V_{bb}\rangle_L^{[N]}/N$. With the help of Eq. (\ref{EEpp}) the energy $\epsilon_L^{ee}(N)$ can be further rewritten as

\be
\epsilon_L^{ee}(N)=\epsilon_L^{Hartree}(N)+
\epsilon_L^{ex-corr}(N),
\label{totalenergy}
\ee
where the Hartree term 
\be
\epsilon_L^{Hartree}(N)=\frac \pi{2N}
\sqrt{\frac \Phi 2} \times\sum_{i=1}^N\sum_{j=1(\neq i)}^N
\tilde V_{iijj}^{LL} - \frac 83\sqrt{\frac N\pi}
\label{hartree}
\ee
includes the energy of the positive background, the exchange-correlation energy

\ba
&&\epsilon_L^{ex-corr}(N)=-
\frac \pi{2N}\sqrt{\frac \Phi 2}
\sum_{i=1}^N\sum_{j=1(\neq i)}^N\left[
\tilde V_{iijj}^{LL}\left(1-\alpha^{ij}_{ij}\right)+
\tilde V_{ijji}^{LL}\alpha^{ij}_{ij}\right]\nonumber \\
&+&
\frac \pi{2N} \sqrt{\frac \Phi 2}\sum_{i=1}^N
\sum_{j=1(\neq i)}^N\sum_{k=1(\neq i,j)}^N(-1)^{j+k}
{\rm sgn}(i-j){\rm sgn}(i-k){\rm Re}\left[2
(\tilde V_{iijk}^{LL}-\tilde V_{ikji}^{LL})\alpha^{ij}_{ik}\right]\nonumber \\
&+&\frac \pi{2N}\sqrt{\frac \Phi 2}
\sum_{i=1}^N\sum_{j=1(\neq i)}^N
\sum_{k=1(\neq i,j)}^N\sum_{l=1(\neq i,j,k)}^N(-1)^{i+j+k+l}
{\rm sgn}(i-k){\rm sgn}(j-l){\rm Re}\left[
\tilde V_{ijkl}^{LL}\alpha^{ik}_{jl}\right]
\label{excorren}
\ea
consists of two-site, three-site and four-site correlations, $\Phi\equiv B/n_s\phi_0=1/\nu$ is the magnetic flux per particle in the units of flux quantum, the matrix elements $\tilde V_{ijkl}^{LL}$ are related with $V_{ijkl}^{LL}$ by the formula

\be
V_{ijkl}^{LL}=\sqrt{\frac\pi 2}\frac{e^2}{\kappa\lambda}\times \tilde V_{ijkl}^{LL}
\label{Vtilde}
\ee
(see Appendix \ref{appVijkl}), and all the energies are measured in units $e^2\sqrt{n_s}/\kappa$. 

The Hartree contribution (\ref{hartree}) can be calculated exactly in the thermodynamic limit $N\to\infty$ (Appendix \ref{app:Hartree}), 

\be
\epsilon_L^{Hartree}=\lim_{N\to\infty}\epsilon_L^{Hartree}(N)=
\pi \sqrt{n_s}
\sum_{\bf G\neq 0}\frac 1G{\cal F}_L^2\left(\frac{(G\lambda)^2}4\right)-
\sqrt{\pi\Phi}\int_0^\infty dx {\cal F}_L^2\left(x^2\right).
\label{hartreefourier}
\ee
Here ${\bf G}$ are reciprocal lattice vectors, the function ${\cal F}_L$ is defined by Eq. (\ref{funccalF}), and the integral is calculated in Eq. (\ref{interg}). The exchange-correlation contribution is calculated, in practice, with a finite number $N$ of lattice points. It converges when $N$ exceeds some number $N_0$, where $N_0=N_0(L,\nu)$ varies from a few tens to a few hundreds, dependent on the overlap of the neighbor single particle wave functions $\chi_L$.

Equations (\ref{eee}), (\ref{hartreefourier}) and (\ref{excorren}), together with (\ref{Viijj})--(\ref{Vijkl}) and (\ref{interg}), provide exact closed-form analytical expressions for the energy of electron-electron interaction in the many-body quantum-mechanical states $\Psi_L$, at all values of the Landau-level filling factor $\nu$, continuously as a function of magnetic field. However complicated, these expressions give a principal possibility to calculate the energy of the system in the thermodynamic limit with any desirable accuracy.

\begin{figure}
\begin{center}
\begin{math}
\epsfxsize=8.5cm
\epsffile{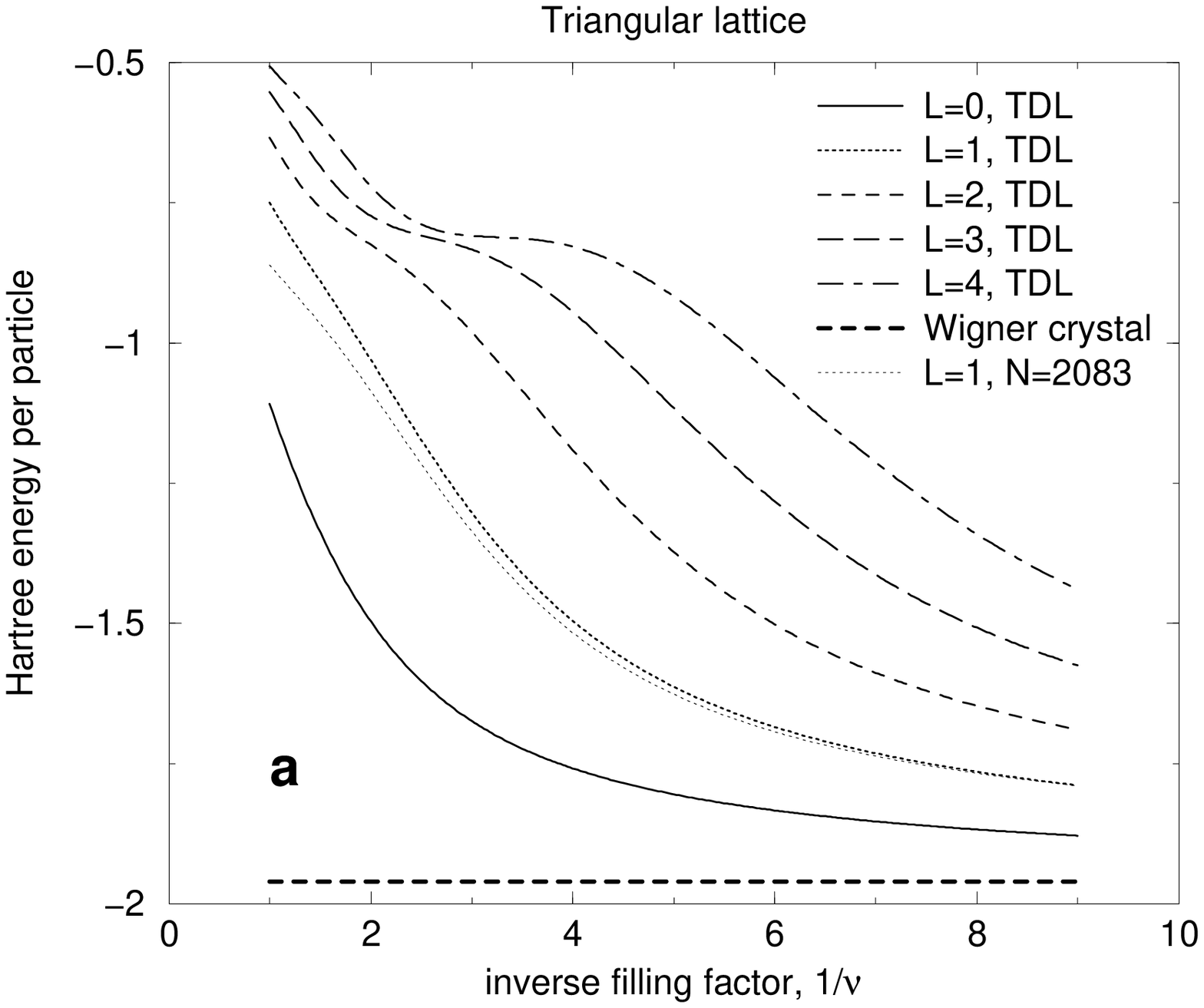}
\end{math}
\begin{math}
\epsfxsize=8.5cm
\epsffile{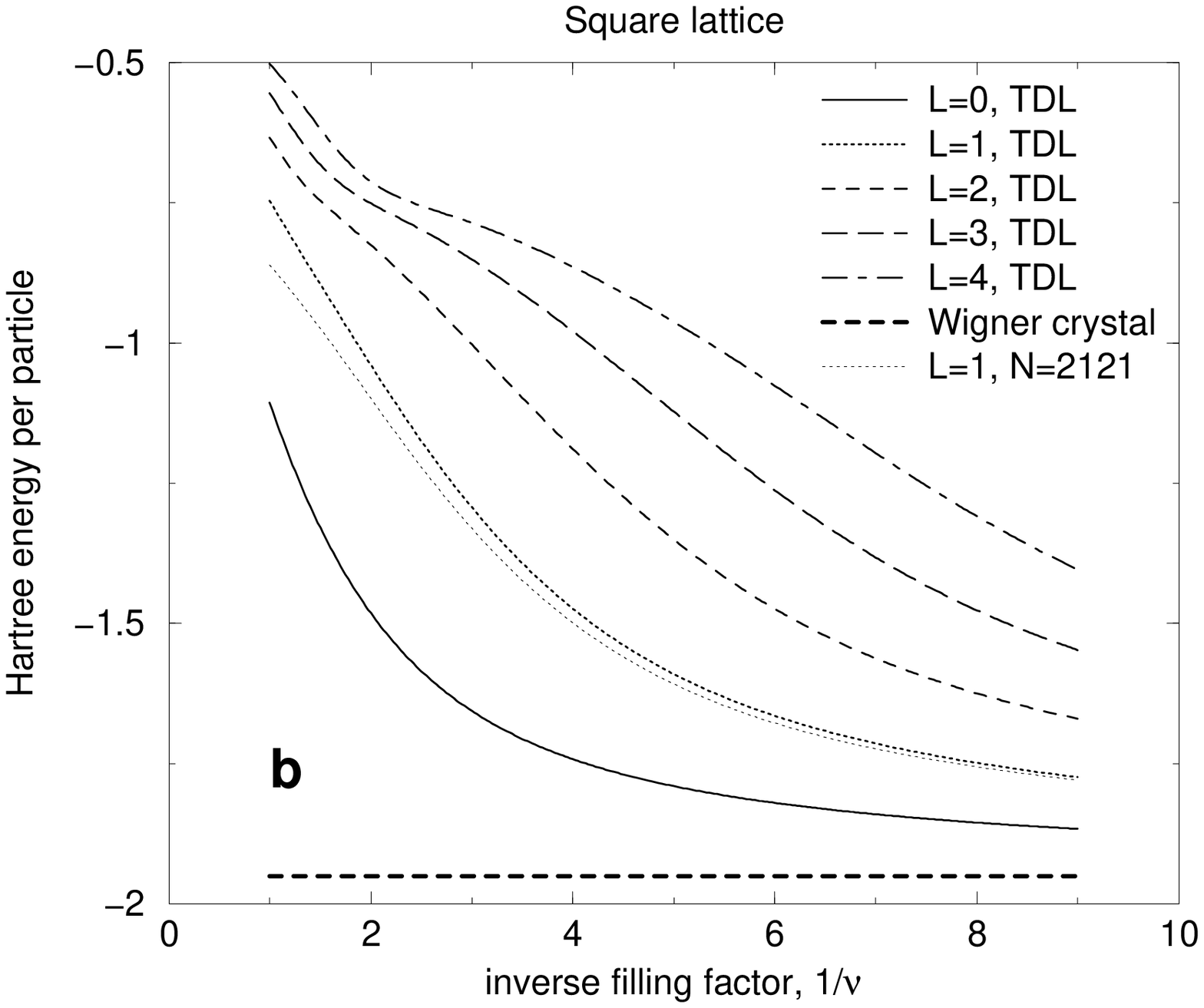}
\end{math}
\end{center}
\caption{Hartree contribution to the energy of the states $\Psi_L$ as a function of the inverse Landau-level filling factor $1/\nu$, calculated in the thermodynamic limit for a few lowest $L$, for a triangular ($a$) and a square ($b$) configurations of lattice points ${\bf R}_j$. The energies of the classical Wigner crystal (\protect\ref{latt-energy}) are shown by thick dashed lines. Thin dotted curves ($L=1$) exhibit the Hartree energies, calculated with a finite number $N>2000$ of lattice points using Eq. (\ref{hartree}). These curves illustrate that the Hartree energy converges very slowly with $N$: to get a few-percent accuracy, one needs to take into account a few thousands particles.}
\label{fig:hartree}
\end{figure}

Figure \ref{fig:hartree} exhibits the magnetic field dependencies of the Hartree contribution to the energy of the states $\Psi_L$, calculated with Eq. (\ref{hartreefourier}) (thick curves), for a few lowest $L$, for the triangular (\ref{tr-lattice}) and the square (\ref{sq-lattice}) lattice of points ${\bf R}_j$ [thin curves at $L=1$ show, for comparison, the results of a finite-$N$ calculations with the formula (\ref{hartree})]. As expected, the Hartree energy increases with $L$, and in the limit of the very strong magnetic field all the curves tend to the classical-Wigner-crystal line. Note that the influence of the quantum-mechanical effects, even in the Hartree approximation, is much stronger than the difference between the energies of the {\it classical} triangular and the square Wigner-crystal lattices, Eq. (\ref{latt-energy}). The {\it quantum-mechanical} ground state may thus be realized by the states (\ref{generalfunction}) or (\ref{generalfunction2}) with {\it different types} of the underlying lattice. Which type of the lattice corresponds to the true quantum-mechanical ground state may also depend on magnetic field. Therefore, different types of the lattice should be tried in the search of the true ground state of the system at different $\nu$.

\begin{figure}
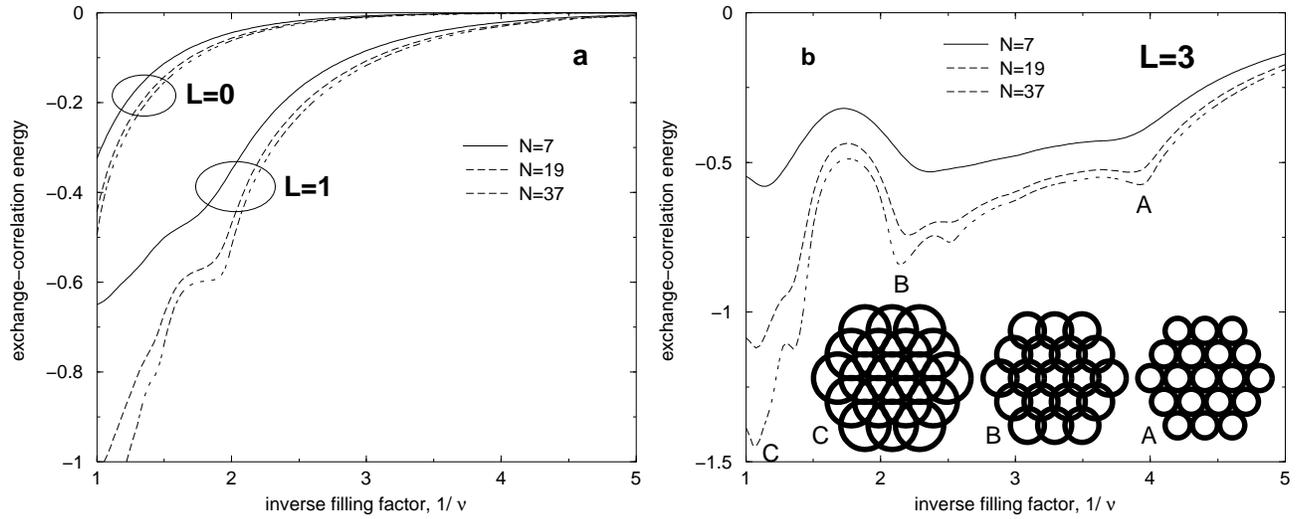

\begin{center}
\begin{math}
\epsfxsize=8.5cm
\epsffile{xcL0-1.epsi}
\end{math}
\begin{math}
\epsfxsize=8.5cm
\epsffile{xcL3.epsi}
\end{math}
\end{center}
\caption{Exchange-correlation contribution to the energy of the states $\Psi_L$ as a function of the inverse Landau-level filling factor $1/\nu$, calculated with a finite number $N$ of (triangular) lattice points $N=7$, 19, and 37, for ($a$) the states $L=0$ and $L=1$, and ($b$) the state $L=3$. Insets in Figure $b$ qualitatively show configurations of electron rings at magnetic fields corresponding to the energy minima in a few characteristic points A, B and C of the $1/\nu$-axis.}
\label{fig:xc}
\end{figure}

\begin{figure}
\begin{center}
\begin{math}
\epsfxsize=8.5cm
\epsffile{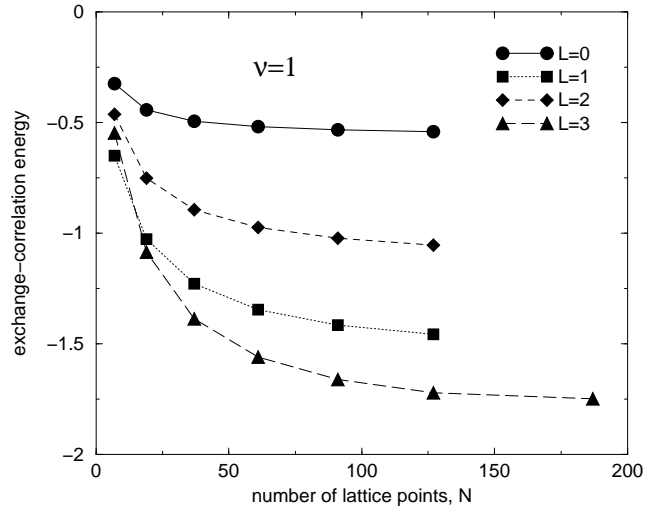}
\end{math}
\end{center}
\caption{Calculated exchange-correlation energy per particle $\epsilon_L^{ex-corr}(N)$, in units $e^2\sqrt{n_s}/\kappa$, at $\nu=1$, as a function of the number of lattice points $N$ in the state $\Psi_{L}$ with $L=0$ to 3. These curves illustrate the convergency of the exchange-correlation contribution to the energy of the states $\Psi_L$: to get a few-percent accuracy one needs to take into account from a few tens to a few hundreds particles, dependent on $L$ and $\nu$, i.e. dependent on the overlap of neighbor single-particle states.}
\label{fig:XCvsN}
\end{figure}

Figure \ref{fig:xc} exhibits the magnetic field dependencies of the exchange-correlation energy (\ref{excorren}) of the states $\Psi_L$, calculated with a finite number $N=7$, 19, and 37 of triangular lattice points, for the states $\Psi_L$ with $L=0$, $1$ and 3. One sees that the energy $\epsilon_L^{ex-corr}(N)$ is negative and grows with $N$ in its absolute value. Figure \ref{fig:XCvsN} shows how the energy $\epsilon_L^{ex-corr}(N)$ converges with $N$. As seen from this Figure, up to a few hundreds particles should be taken into account in order to get a reasonable accuracy of the final result at the considered values of $L$ and $\nu$. If the overlap of the neighbor single-particle states is not negligible, the magnitude of the energy $\epsilon_L^{ex-corr}(N)$ can be comparable with $\epsilon_L^{Hartree}(N)$ (note that the number of Hartree terms grows as $N^2$, while that of the three-site and four-site terms -- as $N^3$ and $N^4$ respectively). The magnetic field dependence of the exchange-correlation energy has a complicated oscillating character, with pronounced minima, resulting from the interference effects of electron waves and the overlap of electron-ring states $\chi_L({\bf r},{\bf R}_j)$. Qualitatively this can be seen in Figure \ref{fig:xc}b, where the insets show the overlap of electron rings at a few characteristic points of the magnetic field axis. The points A and C, for instance, correspond to the touch of the nearest-neighbor and the next-nearest-neighbor rings, respectively. In the point B any three nearest-neighbor rings intersect in one point. The depth of the minima, and their position on the $B$-axis, depend on the index $L$ and the symmetry of the lattice. 

\begin{figure}
\begin{center}
\begin{math}
\epsfxsize=8.5cm
\epsffile{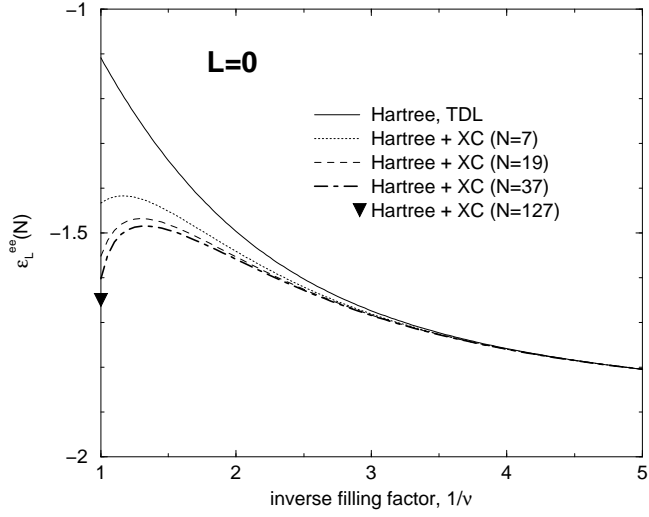}
\end{math}
\end{center}
\caption{Calculated electron-electron interaction energy per particle $\epsilon_L^{ee}(N)=\epsilon_L^{Hartree}+\epsilon_L^{ex-corr}(N)$, in units $e^2\sqrt{n_s}/\kappa$, as a function of $B/n_s\phi_0\equiv 1/\nu$ in the state $\Psi_{L=0}$ (the Wigner crystal, Ref. \protect\cite{Maki83}). The Hartree contribution (the uppermost curve) is calculated exactly in the thermodynamic limit, the exchange-correlation energy -- with a finite number $N=7$, 19, 37, $\dots$ of the (triangular) lattice points.}
\label{fig:L=0}
\end{figure}

\begin{figure}
\begin{center}
\begin{math}
\epsfxsize=8.5cm
\epsffile{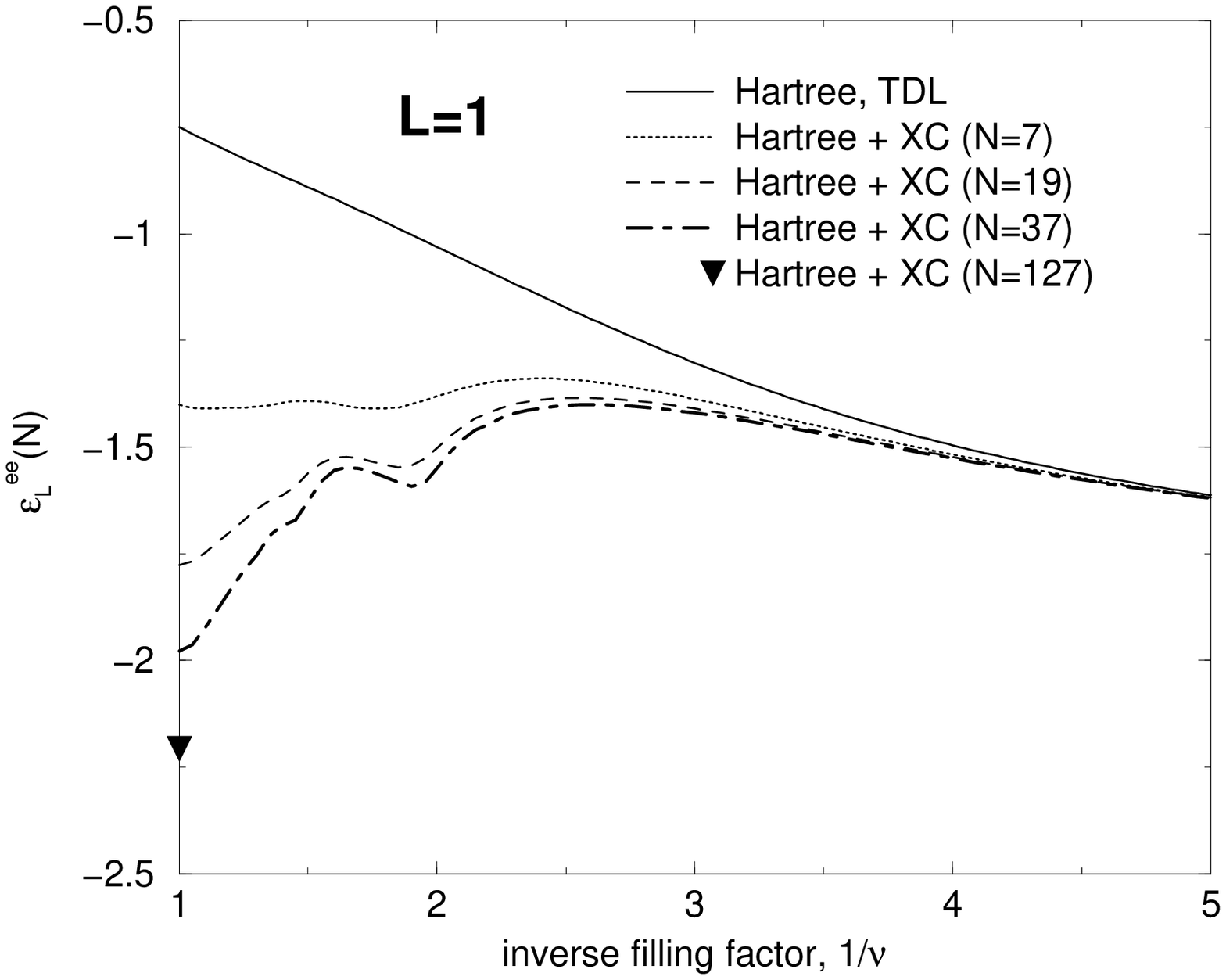}
\end{math}
\end{center}
\caption{The same as in Figure \protect\ref{fig:L=0}, but for the state $\Psi_{L=1}$. }
\label{fig:L=1}
\end{figure}

\begin{figure}
\begin{center}
\begin{math}
\epsfxsize=8.5cm
\epsffile{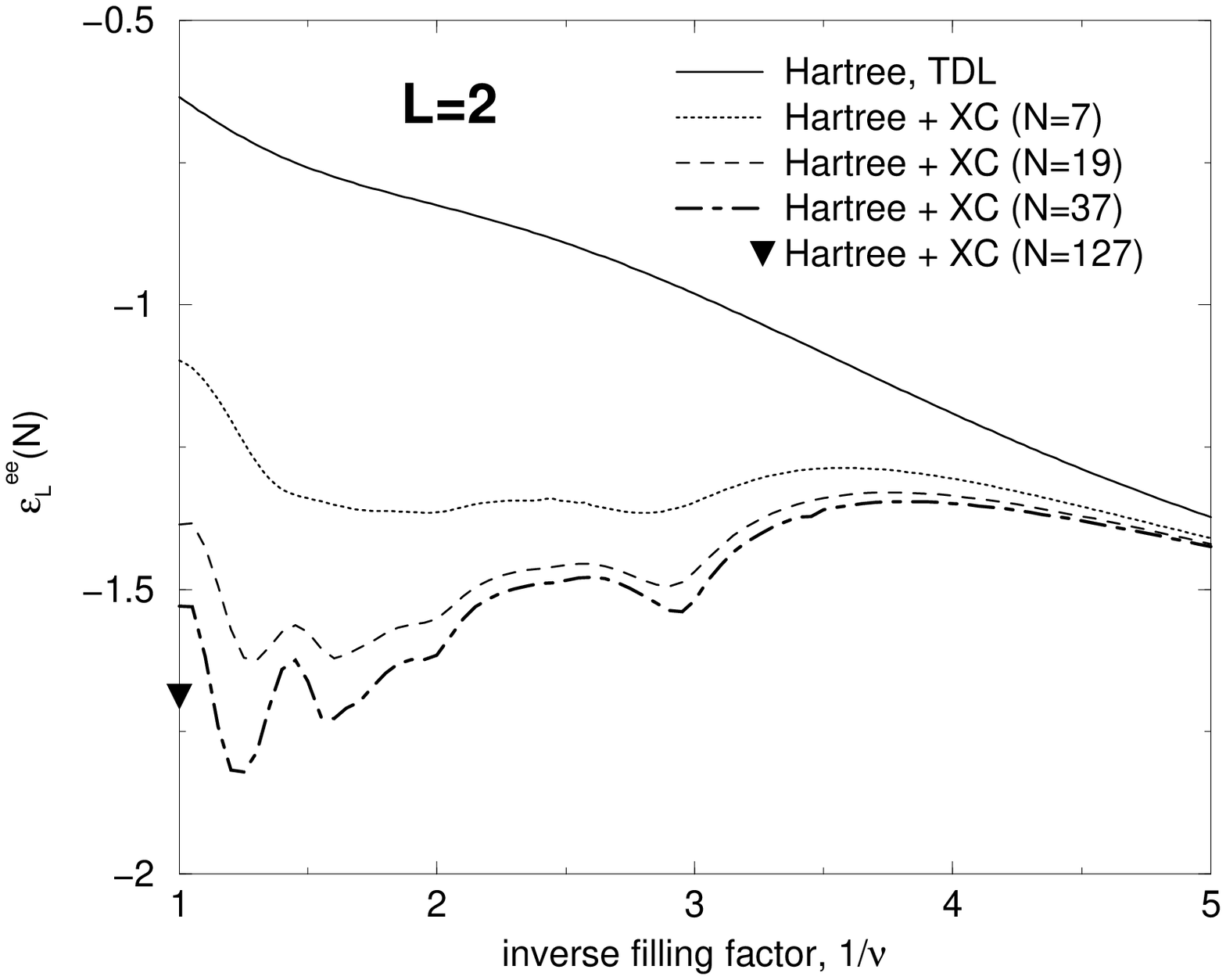}
\end{math}
\end{center}
\caption{The same as in Figure \protect\ref{fig:L=0}, but for the state $\Psi_{L=2}$. }
\label{fig:L=2}
\end{figure}

\begin{figure}
\begin{center}
\begin{math}
\epsfxsize=8.5cm
\epsffile{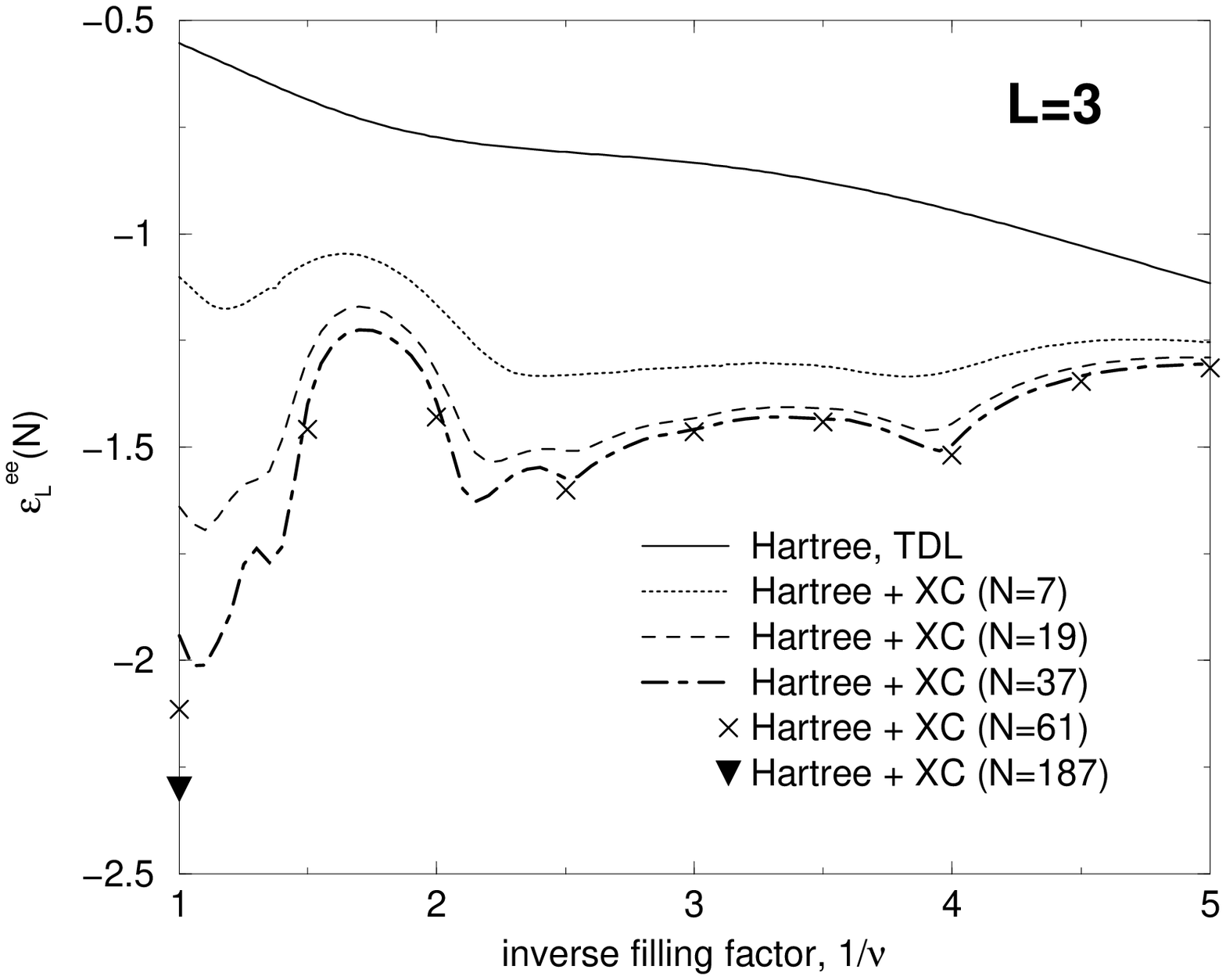}
\end{math}
\end{center}
\caption{The same as in Figure \protect\ref{fig:L=0}, but for the state $\Psi_{L=3}$. }
\label{fig:L=3}
\end{figure}

Figures \ref{fig:L=0}--\ref{fig:L=3} show the total energy of electron-electron interaction $\epsilon_L^{ee}(N)=\epsilon_L^{Hartree}+\epsilon_L^{ex-corr}(N)$ in the states $\Psi_L$ with $L=0$ to 3, as a function of $\Phi=B/n_s\phi_0=1/\nu$. The Hartree contribution (the uppermost curve) has been calculated exactly in the thermodynamic limit, the exchange-correlation contribution -- with a finite number $N$ of lattice points. The total energy $\epsilon_L^{ee}$ also oscillates with magnetic field, due to the physical effects discussed above. As a consequence, the ground state of the system may be realized by the states with different $L$ (and different symmetry of the underlying lattice), in different intervals of magnetic field.

Results of a more complete analysis of the energy and other two-particle properties of the states $\Psi_L$ will be reported in subsequent publications.

\section{Conclusive remarks}
\label{sec:disc}

I have presented basic principles of a new method in the many-body theory of quantum-Hall systems. Only a small part of specific results (for the density and the electron-electron interaction energy of the states $\Psi_{L}$) have been shown with illustrative purposes. A complete study of all properties of the system in different quantum-mechanical states, and all possible consequences from this method will require substantially more time, place and efforts. 

The new approach presented in this paper, involves two essential constituents: (i) a physically clear and transparent way to construct an {\it infinite number} of trial many-body wave functions, {\it for all} magnetic fields $B$ (Section \ref{sec:wavefunc}), and (ii) an exact analytical method of calculating {\it any physical property} of the proposed many-body states (Section \ref{sec:properties}). With the help of this method, one can get comprehensive information on all physical properties of the ground, as well as excited, states of the system at different values of magnetic field. 

The new approach has a huge variational freedom and opens up new fields of research in the many-body theory of quantum-Hall and other many-body systems:

1. It can be used for a systematic search of the ground state of a 2DES at all $\nu$, both in the fractional and in the integer quantum-Hall regimes (the generalization to the case of non-spin-polarized systems and to the region of higher Landau levels has been briefly discussed in Section \ref{variational}). 

2. It makes it possible to study the interplay between effects of disorder and electron-electron correlations in both the integer and the fractional quantum Hall effects (Section \ref{variational}). 

3. It offers an opportunity to separately analyze the Hartree, exchange, and correlation contributions to the total energy of the system at all magnetic fields (Section \ref{sec:twoparticle}). This may find applications in the density functional theory.

4. The new approach can be generalized to and applied in other many-body problems, such as, for instance, many-electrons quantum dots, quasi-one-dimensional electron systems, 2D electrons in inhomogeneous external (electric and magnetic) fields, metal-insulator transitions in 2DES, and other. 

\acknowledgements
This work was done during my stay at the Max-Planck Institute for the Physics of complex Systems, Dresden, and at the Institute for Theoretical Physics, University of Regensburg (support from the Max-Planck Society, the Graduiertenkolleg {\sl Komplexit\"at in Festk\"orpern}, Uni Regensburg, and the DFG-Sonderforschungsbereich 348 is gratefully acknowledged). I thank Ulrich R\"ossler, Allan MacDonald, Peter Fulde, Nadejda Savostianova, Igor Krasovsky, Vladimir Volkov, Vladimir Sandomirsky, Oleg Pankratov, Pawel Hawrylak, Nobuki Maeda, Alexander Rauh and Vladimir Sablikov for helpful discussions, as well as Min-Fong Yang, Robert Laughlin and Francisco Claro for useful comments.

\appendix

\section{Many-body Matrix elements}
\label{app:manybody}

Let $\psi({\bf r},\alpha)$ and $\phi({\bf r},\beta)$ be arbitrary functions
 dependent on the 2D coordinate ${\bf r}$ and arbitrary (maybe, vector)
parameters $\alpha$ and $\beta$ respectively. Let
\be
\Psi_N({\bf r}_1,{\bf r}_2,\dots,{\bf r}_N; 
\alpha_1,\alpha_2,\dots,\alpha_N)=\frac{1}{\sqrt{N!}}\det_N
\left | \psi({\bf r}_i,\alpha_j)\right |
\label{Psi-func}
\ee
and
\be
\Phi_N({\bf r}_1,{\bf r}_2,\dots,{\bf r}_N; 
\beta_1,\beta_2,\dots,\beta_N)=\frac{1}{\sqrt{N!}}\det_N
\left | \phi({\bf r}_i,\beta_j) \right |
\label{Phi-func}
\ee
are $N\times N$ Slater determinants formed from these functions, and $\alpha_i$, $\beta_i$, $i=1,2,\dots,N$, are two sets of $N$ different parameters ($\alpha_i\neq\alpha_j$, $\beta_i\neq\beta_j$). The following theorems can be straightforwardly proved. 

{\bf Theorem 1.} The scalar product

\be
\langle\Psi_N|\Phi_N\rangle\equiv
\int\Psi_N^* 
({\bf r}_1,{\bf r}_2,\dots,{\bf r}_N; 
\alpha_1,\alpha_2,\dots,\alpha_N)\Phi_N({\bf r}_1,{\bf r}_2,\dots,{\bf r}_N; 
\beta_1,\beta_2,\dots,\beta_N)d{\bf r}_1d{\bf r}_2\dots d{\bf r}_N
\ee
is determined by the formula

\be
\langle\Psi_N|\Phi_N\rangle=\det_N {\cal S}_N \equiv \det_N \left | S(\alpha_i,\beta_j) \right |,
\label{S-func}
\ee
where ${\cal S}_N$ is the $N\times N$ matrix with the elements
\be
S(\alpha_i,\beta_j)\equiv S_{ij}=\int\psi^*({\bf r},\alpha_i)\phi({\bf r},\beta_j)d{\bf r}.
\label{S-melement}
\ee

{\bf Theorem 2.} Let $\hat O_1$ be an arbitrary single-particle operator (\ref{O1}). The matrix element $\langle\Psi_N|\hat O_1|\Phi_N\rangle$ is determined by the formula
\be
\langle\Psi_N|\hat O_1|\Phi_N\rangle=
\sum_{i=1}^N\sum_{j=1}^N(-1)^{i+j}o_1(\alpha_i,\beta_j)\det_{N-1}| {\cal S}_{N-1}|^{row\neq i}_{column\neq j},
\label{Hone}
\ee
where the $(N-1)\times (N-1)$ matrix ${\cal S}_{N-1}|^{row\neq i}_{column\neq j}$ is formed from the
matrix ${\cal S}_N$, Eq. (\ref{S-melement}), by deleting the row $i$ and the column $j$, and
\be
o_1(\alpha_i,\beta_j)=\int\psi^*({\bf r},\alpha_i) \hat o_1({\bf r})\phi({\bf r},\beta_j)d{\bf r}.
\label{HoneSP}
\ee

{\bf Theorem 3.} Let $\hat O_2$ be an arbitrary two-particle operator (\ref{O2}). The matrix element $\langle\Psi_N|\hat O_2|\Phi_N\rangle$ is determined by the formula
\be
\langle\Psi_N|\hat O_2|\Phi_N\rangle=
\sum_{i=1}^N\sum_{j=1}^N
\sum_{k=1}^N\sum_{l=1}^N(-1)^{i+j+k+l}
{\rm sgn}(i-k){\rm sgn}(j-l)
o_2(\alpha_i,\alpha_k,\beta_j,\beta_l)\det_{N-2}|
 {\cal S}_{N-2}|^{row\neq i,k}_{column\neq j,l},
\label{Htwo}
\ee
where the $(N-2)\times (N-2)$ matrix ${\cal S}_{N-2}|^{row\neq i,k}_{column\neq j,l}$ is formed from the matrix ${\cal S}_N$, Eq. (\ref{S-melement}), by deleting the rows $i,k$ and the columns $j,l$, the function ${\rm sgn}(n)$ is defined so that ${\rm sgn}(0)=0$, and ${\rm sgn}(n)=\pm 1$ for $n>0$ ($n<0$), and 

\be
o_2(\alpha_i,\alpha_k,\beta_j,\beta_l)=\int\psi^*({\bf r}_1,\alpha_i)
\psi^*({\bf r}_2,\alpha_k)
\hat o_2({\bf r}_1,{\bf r}_2)\phi({\bf r}_1,\beta_j)
\phi({\bf r}_2,\beta_l)d{\bf r}_1d{\bf r}_2.
\label{2partme}
\ee

\section{Coulomb matrix elements}
\label{appVijkl}

In this Section I present results of calculation of the Coulomb matrix elements 
\be
V_{ijkl}^{LL^\prime}=\int\int d{\bf r}_ad{\bf r}_b (\chi_{ai}^L)^*\chi_{aj}^{L^\prime}\frac {e^2}{\kappa|{\bf r}_a-{\bf r}_b|}(\chi_{bk}^L)^*\chi_{bl}^{L^\prime},
\label{Vijkl_LLprime}
\ee
where, for generality, I assume that the indexes $L$ and $L^\prime$ are not necessarily equal to each other. In all the formulas below $\xi_i$ are defined by Eq. (\ref{xi-define}), $\eta_{ij}=\xi_i-\xi_j$, and $F^{(n)}(z)$ is the $n$-th derivative of the function 
\be
F(z)=e^{-z}I_0(z),
\label{functionF}
\ee
where $I_0$ is the modified Bessel function. 

There exist five different types of the integrals $V_{ijkl}^{LL^\prime}$, dependent on whether one or two pairs of subscripts equal to each other, or all the indexes $ijkl$ are different:

1. Direct Coulomb two-site matrix elements $V_{iijj}^{LL^\prime}$,

\ba
V_{iijj}^{LL^\prime}&=&\sqrt{\frac\pi 2}\frac{e^2}{\kappa\lambda}
\times\frac{1} {L!L^\prime!2^{L+L^\prime}}
\sum_{p=0}^L \frac{(-1)^p L!}{p!(L-p)!} \left(\frac{\eta^2}8\right)^p
\nonumber \\
&\times&
\sum_{q=0}^{L^\prime} 
\frac{(-1)^q L^\prime!\delta_{L-p,L^\prime-q}
(L-p+L^\prime-q)!}{q!(L^\prime-q)!}\left(\frac{(\eta^*)^2}8\right)^q
\sum_{i=0}^{2p}
\frac{(2p)!(-2)^i}{i!(2p-i)!}
\nonumber \\
&\times&
\sum_{k=0}^{\min\{2q,2p-i\}}\frac{(2q)!(2p-i)!}{k!(2q-k)!(2p-i-k)!}
\frac 1{\zeta^k}
\sum_{j=0}^{2p-i}\frac{(2p-i)!2^j}{j!(2p-i-j)!}
F^{(2p+2q-i-j-k)}(\zeta).
\label{Viijj}
\ea
Here and in Eq. (\ref{Vijji}) below, $\zeta=\eta\eta^*/4$, and $\eta=\eta_{ij}$.

2. Exchange two-site matrix elements $V_{ijji}^{LL^\prime}$,

\ba
V_{ijji}^{LL^\prime}&=&\sqrt{\frac\pi 2}\frac{e^2}{\kappa\lambda}
\times\frac{\exp(-\eta\eta^*/2)} {L!L^\prime!2^{L+L^\prime}}
\sum_{p=0}^L \frac{(-1)^p L!}{p!(L-p)!}\left(\frac{\eta^2} 8\right)^{p} 
\nonumber \\
&\times&
\sum_{q=0}^{L^\prime} 
\frac{(-1)^q L^\prime!\delta_{L-p,L^\prime-q}
(L-p+L^\prime-q)!}{q!(L^\prime-q)!}\left(\frac{(\eta^*)^2} 8\right)^{q}
\nonumber \\
&\times&
\sum_{i=0}^{2p}
\frac{(2p)!(-2)^i}{i!(2p-i)!}
\sum_{j=0}^{2q}
\frac{(2q)!(-2)^j}{j!(2q-j)!}
\nonumber \\
&\times&
\sum_{k=0}^{\min\{2p-i,2q-j\}}\frac{(2p-i)!(2q-j)!}{k!(2p-i-k)!(2q-j-k)!}
\frac 1{\zeta^k}F^{(2p+2q-i-j-k)}(\zeta).
\label{Vijji}
\ea

3. Direct three-site matrix elements  $V_{iikl}^{LL^\prime}$,

\ba
V_{iikl}^{LL^\prime}&=&\sqrt{\frac\pi 2}\frac{e^2}{\kappa\lambda}
\times\frac{(\eta_{ik}\eta_{il})^L(\eta_{il}^*\eta_{ik}^*)^{L^\prime}
}{L!L^\prime!(-2)^{L+L^\prime}} 
\exp\left(-\frac{\eta_{il}\eta_{il}^*}2+\frac{\xi_i\eta_{kl}^*-\xi_k^*\eta_{kl}}2\right)
\nonumber \\
&\times&
\sum_{p=0}^L \frac{(-1)^p L!}{p!(L-p)!} \left(
\frac {\eta_{il}}{2\eta_{ik}} \right)^p
\sum_{q=0}^{L^\prime} \frac{(-1)^q L^\prime!}{q!(L^\prime-q)!}\left(
\frac {\eta_{ik}^*}{2\eta_{il}^*}\right)^q
\nonumber \\
&\times&
\sum_{r=0}^{\min\{L+p, L^\prime+q\}}
\frac{(L+p)! (L^\prime+q)!}{r!(L+p-r)! (L^\prime+q-r)!}
\left( \frac 2{\eta_{il}\eta_{ik}^*} \right)^r
\nonumber \\
&\times&
\sum_{i=0}^{L-p} \frac{ (L-p)!}{i!(L-p-i)!}
\left(\frac {\eta_{il}}{2\eta_{ik}}\right)^{i}
\sum_{j=0}^{L^\prime-q} \frac{ (L^\prime-q)!}{j!(L^\prime-q-j)!}
\left(\frac {\eta_{ik}^*}{2\eta_{il}^*}\right)^{j}
\sum_{k=0}^{L+p-r} \frac{ (L+p-r)!}{k!(L+p-r-k)!2^k}
\nonumber \\
&\times&
\sum_{l=0}^{L^\prime+q-r} \frac{ (L^\prime+q-r)!}{l!(L^\prime+q-r-l)!2^l}
\sum_{s=0}^{\min\{i+k,j+l\}}
\frac{(i+k)!(j+l)!}{s!(i+k-s)!(j+l-s)!}
\left(-\frac4{\eta_{il}\eta_{ik}^*}\right)^s
\nonumber \\
&\times&
F^{(i+j+k+l-s)}\left(-\frac{\eta_{il}\eta_{ik}^*}4\right).
\label{Viikl}
\ea

4. Exchange three-site matrix elements $V_{ijki}^{LL^\prime}$,

\ba
V_{ijki}^{LL^\prime}&=&\sqrt{\frac{\pi}2}\frac{e^2}{\kappa\lambda}\times
\exp\left(-\frac{\xi_j\xi_j^*+\xi_k\xi_k^*}2+\xi_k^*\xi_j-
\frac{\eta_{ij}\eta_{ik}^*}2\right)\frac 
{(\eta_{ij}\eta_{ik})^L(\eta_{ij}^*\eta_{ik}^*)^{L^\prime}}
{L!L^\prime!4^{L+L^\prime}}
\nonumber \\
&\times&
\sum_{i=0}^{L}\frac {L!}{i!(L-i)!}(-2)^i
\sum_{j=0}^{L^\prime}\frac {L^\prime!}{j!(L^\prime-j)!}(-2)^j
\sum_{k=0}^{L}\frac {L!}{k!(L-k)!}\left(-\frac{\eta_{ij}}{2\eta_{ik}}\right)^k
\nonumber \\
&\times&
\sum_{l=0}^{L^\prime}\frac {L^\prime!}{l!(L^\prime-l)!}
\left(-\frac{\eta_{ik}^*}{2\eta_{ij}^*}\right)^l
\sum_{s=0}^{\min\{i+k,j+l\}}
\frac{(i+k)!(j+l)!}{s!(i+k-s)!(j+l-s)!}
\left(\frac 2{\eta_{ij}\eta_{ik}^*}\right)^s
\nonumber \\
&\times&
\sum_{p=0}^{L-k}\frac {(L-k)!}{p!(L-k-p)!}
\left(-\frac{\eta_{ij}}{4\eta_{ik}}\right)^p
\sum_{q=0}^{L^\prime-l}\frac{(L^\prime-l)!}{q!(L^\prime-l-q)!}
\left(-\frac{\eta_{ik}^*}{4\eta_{ij}^*}\right)^q
\nonumber \\
&\times&
\sum_{r=0}^{\min\{L-i+p,L^\prime-j+q\}}
\frac{(L-i+p)!(L^\prime-j+q)!}{r!(L-i+p-r)!(L^\prime-j+q-r)!}
\left(\frac 4{\eta_{ij}\eta_{ik}^*}\right)^r
\nonumber \\
&\times&
F^{(L^\prime-j+q+L-i+p-r)}\left(\frac{\eta_{ij}\eta_{ik}^*}4\right).
\label{Vijki}
\ea

5. Four-site matrix elements $V_{ijkl}^{LL^\prime}$,

\ba
V_{ijkl}^{LL^\prime}&=&
\frac{e^2}{\kappa\lambda}\sqrt{\frac\pi 2}\times
\frac {(\eta_{ij}\eta_{kl})^L (\eta_{ij}^*\eta_{kl}^*)^{L^\prime}} 
{L!L^\prime!}
\exp\left\{-\frac{\xi_i\xi_i^*+\xi_j\xi_j^*+\xi_k\xi_k^*+
\xi_l\xi_l^*}2+\xi_j\xi_i^*+\xi_l\xi_k^*\right\} 
\nonumber \\
&\times&
\sum_{i=0}^L\frac{L!}{i!(L-i)!}\left(-\frac {\eta_{jl}}{4\eta_{ij}}\right)^i
\sum_{k=0}^L\frac{L!}{k!(L-k)!}\left(\frac {\eta_{jl}}{4\eta_{kl}}\right)^k
\sum_{j=0}^{L^\prime}\frac{L^\prime!}{j!(L^\prime-j)!}
\left(\frac {\eta_{ik}^*}{4\eta_{ij}^*}\right)^j
\nonumber \\
&\times&
\sum_{l=0}^{L^\prime}\frac{L^\prime!}{l!(L^\prime-l)!}
\left(-\frac {\eta_{ik}^*}{4\eta_{kl}^*}\right)^l
\sum_{p=0}^k\frac{k!}{p!(k-p)!}
\left(\frac 4{\eta_{jl}\eta_{ik}^*}\right)^p
\nonumber \\
&\times&
\sum_{q=\max\{0,p-j-l\}}^{\min\{p,L^\prime-l\}}
\frac{p!(L^\prime-l)!(j+l)!}
{q!(p-q)!(L^\prime-l-q)!(j+l-p+q)!}
\left(-\frac {\eta_{ik}^*}{\eta_{kl}^*}\right)^q
\sum_{r=0}^i\frac{i!}{r!(i-r)!}\left(\frac 4{\eta_{jl}\eta_{ik}^*}\right)^r
\nonumber \\
&\times&
\sum_{s=\max\{0,p+r-j-l-q\}}^{\min\{r,L^\prime-j\}}
\frac{r!(L^\prime-j)!(j+l-p+q)!}{s!(r-s)!(L^\prime-j-s)!(j+l-p+q-r+s)!}
\left(\frac {\eta_{ik}^*}{\eta_{ij}^*}\right)^s
\nonumber \\
&\times&
F^{(i+j+k+l-p-r)}(\eta_{jl}\eta_{ik}^*/4).
\label{Vijkl}
\ea

\section{Hartree approximation}
\label{app:Hartree}

The Hartree approximation works well when the overlap of the neighbor single-particle wave functions is negligible. In this approximation the many-body wave function corresponding to (\ref{PsiL}) is written as 

\be
\Psi_{L,Hartree}^{[N]}({\bf r}_1,{\bf r}_2,\dots,{\bf r}_N)=\chi_{11}^L\chi_{22}^L\dots \chi_{NN}^L.
\ee
The functions $\chi$ are normalized. Hence, $\langle \Psi_{L,Hartree}^{[N]}| \Psi_{L,Hartree}^{[N]}\rangle =1$, and 
\be
\langle \Psi_{L,Hartree}^{[N]} | V_{ee}|\Psi_{L,Hartree}^{[N]} \rangle =
\frac
12\sum_{i=1}^N\sum_{j=1,j\neq i}^N V_{iijj}^{LL},
\ee
where $V_{iijj}^{LL}$ is given by (\ref{Viijj}).

The density of electrons in the state $\Psi_{L,Hartree}$ is

\be
n_e^{L,Hartree}({\bf r})=\lim_{N\to\infty}\sum_{i=1}^N
|\psi_L({\bf r}-{\bf R}_i)|^2\equiv \sum_{\bf G}n_e^{L,Hartree}({\bf G})e^{i{\bf G}\cdot{\bf r}},
\label{Hartreedensity}
\ee
where ${\bf G}$ are reciprocal lattice vectors, 
\ba
n_e^{L,Hartree}({\bf G})&=&n_s\int_{cell}\lim_{N\to\infty}\sum_{i=1}^N
|\psi_L({\bf r}-{\bf R}_i)|^2e^{-i{\bf G}\cdot {\bf r}}d{\bf r}\nonumber \\
&=&
n_s\int|\psi_L({\bf r})|^2e^{-i{\bf G}\cdot{\bf r}}d{\bf r}=n_s{\cal F}_L\left((G\lambda)^2/4\right),
\ea
and 
\be
{\cal F}_L(x)=e^{-x}L_L(x).
\label{funccalF}
\ee

The formula (\ref{hartree}) for the Hartree energy can be transformed as follows

\ba
&&N\frac{e^2\sqrt{n_s}}\kappa \epsilon_L^{Hartree}(N)=\frac 1{2}
\sum_{i=1}^N\sum_{j=1(\neq i)}^N
V_{iijj}^{LL}-\frac{e^2}{2\kappa}\int\int\frac{d{\bf r}_ad{\bf r}_b}{|{\bf r}_a-{\bf r}_b|}n_b({\bf r}_a)n_b({\bf r}_b)\nonumber \\
&=&
\frac{e^2}{2\kappa}\int\int\frac{d{\bf r}_ad{\bf r}_b}{|{\bf r}_a-{\bf r}_b|}
\left\{n_e^{L,Hartree}({\bf r}_a)n_e^{L,Hartree}({\bf r}_b)
-n_b({\bf r}_a)n_b({\bf r}_b)-
N
|\psi_L({\bf r}_a)|^2
|\psi_L({\bf r}_b)|^2\right\}
\label{hartree1}
\ea
where $n_b({\bf r})=n_s\theta(R_b-r)$ is the density of the positive background. Substituting here the Fourier expansion from (\ref{Hartreedensity}) I get Eq. (\ref{hartreefourier}) from the main text of the paper. The integral there can be analytically calculated,
\be
\int_0^\infty dx {\cal F}_L^2\left(x^2\right)=
\frac 12\sqrt{\frac\pi 2}
\sum_{i=0}^L\frac{L!}{i!i!(L-i)!}\left(-\frac 14\right)^i
\sum_{j=0}^L\frac{L!}{j!j!(L-j)!}\left(-\frac 14\right)^j
[2(i+j)-1]!!
\label{interg}
\ee

\begin{thebibliography}{10}

\bibitem[\star]{address} Present address: Theoretical Physics II, Institute for Physics, University of Augsburg, D-86135 Augsburg, Germany; e-mail: sergey.mikhailov@physik.uni-augsburg.de

\bibitem{Fukuyama79}
H. Fukuyama, P.~M. Platzman, and P.~W. Anderson, Phys. Rev. B 19
  (1979)  5211.

\bibitem{Yoshioka79}
D. Yoshioka and H. Fukuyama, J. Phys. Soc. Japan  47  (1979)  394.

\bibitem{Bychkov81}
Y.~A. Bychkov, S.~V. Iordanskii, and G.~M. Eliashberg, JETP Lett.  33
   (1981) 143.

\bibitem{Yoshioka83b}
D. Yoshioka and P.~A. Lee, Phys. Rev. B  27   (1983) 4986.

\bibitem{Maki83}
K. Maki and X. Zotos, Phys. Rev. B  28   (1983) 4349.

\bibitem{Yoshioka83a}
D. Yoshioka, B.~I. Halperin, and P.~A. Lee, Phys. Rev. Lett.  50 
  (1983) 1219.

\bibitem{Laughlin83}
R.~B. Laughlin, Phys. Rev. Lett.  50   (1983) 1395.

\bibitem{Haldane83}
F.~D.~M. Haldane, Phys. Rev. Lett.  51   (1983) 605.

\bibitem{Yoshioka84b}
D. Yoshioka, Phys. Rev. B  29   (1984) 6833.

\bibitem{Lam84}
P.~K. Lam and S.~M. Girvin, Phys. Rev. B  30   (1984) 473.

\bibitem{Levesque84}
D. Levesque, J.~J. Weis, and A.~H. MacDonald, Phys. Rev. B  30 
  (1984) 1056.

\bibitem{Haldane85}
F.~D.~M. Haldane and E.~H. Rezayi, Phys. Rev. Lett.  54   (1985) 237.

\bibitem{MacDonald85}
A.~H. MacDonald, G.~C. Aers, and M.~W.~C. Dharma-wardana, Phys. Rev. B 
  31   (1985) 5529.

\bibitem{Kivelson86}
S. Kivelson, C. Kallin, D.~P. Arovas, and J.~R. Schrieffer, Phys. Rev. Lett.
   56   (1986) 873.

\bibitem{Morf86}
R. Morf and B.~I. Halperin, Phys. Rev. B  33   (1986) 2221.

\bibitem{Jain89a}
J.~K. Jain, Phys. Rev. Lett.  63   (1989) 199.

\bibitem{Halperin93}
B.~I. Halperin, P.~A. Lee, and N. Read, Phys. Rev. B  47   (1993) 7312.

\bibitem{Aleiner95}
I.~L. Aleiner and L.~I. Glazman, Phys. Rev. B  52   (1995) 11296.

\bibitem{Koulakov96}
A.~A. Koulakov, M.~M. Fogler, and B.~I. Shklovskii, Phys. Rev. Lett.  76
   (1996) 499.

\bibitem{Fogler96}
M.~M. Fogler, A.~A. Koulakov, and B.~I. Shklovskii, Phys. Rev. B  54
   (1996) 1853.

\bibitem{Moessner96}
R. Moessner and J.~T. Chalker, Phys. Rev. B  54   (1996) 5006.

\bibitem{Fogler97}
M.~M. Fogler and A.~A. Koulakov, Phys. Rev. B  55   (1997) 9326.

\bibitem{Jungwirth00}
T. Jungwirth, A.~H. MacDonald, L. Smrcka, and S.~M. Girvin, Physica E  6
   (2000) 43.

\bibitem{MacDonald00}
A.~H. MacDonald and M.~P.~A. Fisher, Phys. Rev. B  61   (2000) 5724.

\bibitem{Klitzing80}
K. von Klitzing, G. Dorda, and M. Pepper, Phys. Rev. Lett.  45 
  (1980) 494.

\bibitem{Tsui82}
D.~C. Tsui, H.~L. Stormer, and A.~C. Gossard, Phys. Rev. Lett.  48 
  (1982) 1559.

\bibitem{Wigner34}
E.~P. Wigner, Phys. Rev.  46   (1934) 1002.

\bibitem{Lilly99a}
M.~P. Lilly, K.~B. Cooper, J.~P. Eisenstein, L.~N. Pfeiffer, and K.~W. West,
  Phys. Rev. Lett.  82   (1999) 394.

\bibitem{Lilly99b}
M.~P. Lilly, K.~B. Cooper, J.~P. Eisenstein, L.~N. Pfeiffer, and K.~W. West,
  Phys. Rev. Lett.  83   (1999) 824.

\bibitem{Du99}
R.~R. Du, D.~C. Tsui, H.~L. Stormer, L.~N. Pfeiffer, K.~W. Baldwin, and K.~W.
  West, Solid State Commun.  109   (1999) 389.

\bibitem{Pan99}
W. Pan, R.~R. Du, H.~L. Stormer, D.~C. Tsui, L.~N. Pfeiffer, K.~W. Baldwin, and
  K.~W. West, Phys. Rev. Lett.  83   (1999) 820.

\bibitem{Cooper99}
K.~B. Cooper, M.~P. Lilly, J.~P. Eisenstein, L.~N. Pfeiffer, and K.~W. West,
  Phys. Rev. B  60   (1999) R11285.

\bibitem{Du00}
R.~R. Du, W. Pan, H.~L. Stormer, D.~C. Tsui, L.~N. Pfeiffer, K.~W. Baldwin, and
  K.~W. West, Physica E  6   (2000) 36.

\bibitem{Eisenstein00}
J.~P. Eisenstein, M.~P. Lilly, K.~B. Cooper, L.~N. Pfeiffer, and K.~W. West,
  Physica E  6   (2000) 29.

\bibitem{Chaplik72}
A.~V. Chaplik, Zh. Eksp. Teor. Fiz.  62   (1972) 746, [Sov. Phys.-JETP
   35 (1972) 395].

\bibitem{Meissner76}
G. Meissner, H. Namaizawa, and M. Voss, Phys. Rev. B  13   (1976) 1370.

\bibitem{Bonsall77}
L. Bonsall and A.~A. Maradudin, Phys. Rev. B  15   (1977) 1959.

\bibitem{Fock28}
V. Fock, Z. Phys.  47   (1928) 446.

\bibitem{Landau30}
L. Landau, Z. Phys.  64   (1930) 629.

\bibitem{Teller30}
E. Teller, Z. Phys.  67   (1930) 311.

\bibitem{Tosatti83}
E. Tosatti and M. Parrinello, Lett. Nuovo Cimento  36   (1983) 289.

\bibitem{Claro85}
F. Claro, Solid State Commun.  53   (1985) 27.

\bibitem{Claro87}
F. Claro, Phys. Rev. B  35   (1987) 7980.

\bibitem{Brown64}
E. Brown, Phys. Rev.  A133   (1964) 1038.

\bibitem{Dana83}
I. Dana and J. Zak, Phys. Rev. B  28   (1983) 811.

\bibitem{Thouless84}
D.~J. Thouless, J. Phys. C  17   (1984) L325.

\bibitem{Imai90}
N. Imai, K. Ishikawa, T. Matsuyama, and I. Tanaka, Phys. Rev. B  42
   (1990) 10610.

\bibitem{Ishikawa92}
K. Ishikawa, Suppl. Prog. Theor. Phys.  107   (1992) 167.

\bibitem{Ishikawa95}
K. Ishikawa, N. Maeda, and K. Tadaki, Phys. Rev. B  51   (1995) 5048.

\bibitem{Ishikawa98}
K. Ishikawa, N. Maeda, T. Ochiai, and H. Suzuki, Phys. Rev. B  58 
  (1998) 1088.

\bibitem{Ishikawa99}
K. Ishikawa, N. Maeda, T. Ochiai, and H. Suzuki, Physica E  4 
  (1999) 37.

\bibitem{Maeda00}
N. Maeda, Phys. Rev. B  61   (2000) 4766.

\bibitem{Cha94}
M.-C. Cha and H.~A. Fertig, Phys. Rev. B  50   (1994) 14368.

\end{thebibliography}
\end{document}